\documentclass[acmsmall]{acmart}

\AtBeginDocument{%
  \providecommand\BibTeX{{%
    \normalfont B\kern-0.5em{\scshape i\kern-0.25em b}\kern-0.8em\TeX}}}

\setcopyright{acmcopyright}
\copyrightyear{2022}
\acmYear{2022}
\acmDOI{?}

\acmJournal{TOPS}
\acmVolume{00}
\acmNumber{0}
\acmArticle{000}
\acmMonth{1}

\usepackage{amsthm}

\usepackage{fixltx2e}
\usepackage{blindtext}
\usepackage{verbatim}
\usepackage{url}
\usepackage{color}
\usepackage{nicefrac}
\usepackage{graphicx}
\usepackage{multirow}
\usepackage{rotate}
\usepackage{stackengine} 
\usepackage{bm} 
\usepackage{thmtools}
\usepackage{thm-restate}
\usepackage{relsize} 
\usepackage[ruled,noend]{algorithm2e} 
\usepackage{multicol}
\usepackage{wrapfig}
\usepackage{ulem} 
\usepackage{float}
\usepackage{subcaption}
\usepackage{hyperref}
\allowdisplaybreaks[1]

\makeatletter
\newcommand{\figcaption}[1]{\def\@captype{figure}\caption{#1}}
\newcommand{\tblcaption}[1]{\def\@captype{table}\caption{#1}}
\makeatother

\newtheorem{Theorem}{Theorem}

\newtheorem{Definition}[Theorem]{Definition}
\newtheorem{Example}[Theorem]{Example}
\newcommand\Fig[1] {Fig.~\ref{#1}}
\newcommand\Figs[1] {Figs.~\ref{#1}}

\newcommand{\cals}{\mathcal{S}}
\newcommand{\calx}{\mathcal{X}}
\newcommand{\caly}{\mathcal{Y}}
\newcommand{\calz}{\mathcal{Z}}
\newcommand{\calc}{\mathcal{C}}

\newcommand{\cala}{\mathcal{A}}
\newcommand{\cald}{\mathcal{D}}
\newcommand{\calk}{\mathcal{K}}
\newcommand{\calw}{\mathcal{W}}

\newcommand{\cali}{\mathcal{I}}

\DeclareMathAlphabet{\mathbbm}{U}{bbm}{m}{n}
\newcommand{\reals}{\mathbb{R}}
\newcommand{\distr}{\mathbb{D}}

\newcommand{\payd}{\textit{u}_{\sf d}}
\newcommand{\paya}{\textit{u}_{\sf a}}

\newcommand{\Payd}{\textit{U}_{\sf d}}
\newcommand{\Paya}{\textit{U}_{\sf a}}
\newcommand{\Pay}{\textit{U}}

\newcommand{\eqdef}{\ensuremath{\stackrel{\mathrm{def}}{=}}}
\newcommand{\argmax}{\operatornamewithlimits{argmax}}
\newcommand{\argmin}{\operatornamewithlimits{argmin}}
\newcommand{\supp}[1]{{\sf supp}(#1)}
\newcommand{\expect}{\operatornamewithlimits{\mathbbm{E}}}
\newcommand{\expectDouble}[2]{\operatornamewithlimits{\displaystyle\mathbbm{E}}_{\substack{#1\\ #2}}}

\newcommand{\vf}{\mathbbm{V}} 
\newcommand{\priorvf}[1]{\vf\!\left[#1\right]} 
\newcommand{\postvf}[2]{\vf\!\left[#1,#2\right]} 
\newcommand{\vbayes}{\vf^{\mathrm{Bayes}}} 
\newcommand{\postvbayes}[2]{\vbayes\!\left[#1,#2\right]} 

\newcommand{\vdp}{\mathbbm{V}^{\rm DP}}

\newcommand{\samplefrom}{\leftarrow}

\newcommand{\bigadd}{\operatorname{\sum}}

\newcommand{\conc}{\diamond} 
\newcommand{\bigconc}{\mathop{\mathlarger{\Diamond}}} 
\newcommand{\bigsqcupdot}{\mathop{\ooalign{$\bigsqcup$\crcr$\hss\cdot\hss$}}}
\newcommand{\VChoice}[2]{{\bigsqcupdot_{#1 \samplefrom #2}}}
\newcommand{\VChoiceDouble}[4]{{\bigsqcupdot_{\substack{#1 \samplefrom #2\\ #3 \samplefrom #4}}}}

\def\dist#1{\mathbbm{D}{#1}}

\newcommand{\qm}[1]{``#1''}
\newcommand{\true}{T}
\newcommand{\false}{F}

\newcommand{\smallersum}[1]{\textstyle{\sum_{#1}\:}}

\newcommand{\qif}{\texttt{QIF}}
\newcommand{\edp}{\texttt{DP}}
\newcommand{\RR}[2]{\mathit{RR}_{#1}^{#2}}

\newif\ifcommentson\commentsonfalse

\usepackage{thmtools}
\usepackage{thm-restate}

\usepackage{cleveref}

\DeclareMathVersion{normal2}
\begin{document}

\title{Information Leakage Games: Exploring Information as a Utility Function}

\author{M\'{a}rio S. Alvim}
\authornote{All authors contributed equally to this research,
hence are listed in alphabetical order}
\orcid{0000-0002-4196-7467}
\affiliation{%
  \institution{UFMG}
  \streetaddress{Av. Ant\^{o}nio Carlos, 6627}
  \city{Belo Horizonte}
  \country{Brazil}}

\author{Konstantinos Chatzikokolakis}
\orcid{0000-0002-3081-5775}
\affiliation{%
  \institution{University of Athens}
  \streetaddress{15784, Ilisia}
  \city{Athens}
  \country{Greece}}
\affiliation{%
  \institution{CNRS}
  \city{Gif-sur-Yvette}
  \country{France}}

\author{Yusuke Kawamoto}
\orcid{0000-0002-2151-9560}
\affiliation{%
  \institution{AIST}
  \streetaddress{2-4-7 Aomi, Koto-ku}
  \city{Tokyo}
  \country{Japan}
}

\author{Catuscia Palamidessi}
\orcid{0000-0003-4597-7002}
\affiliation{%
  \institution{Inria Saclay}
  \streetaddress{B\^{a}timent Alan Turing, rue Honor\'{e} d'Estienne d'Orves, Campus de l'\'{E}cole Polytechnique}
  \city{Palaiseau}
  \country{France}}
\affiliation{%
  \institution{\'{E}cole Polytechnique}
  \city{Palaiseau}
  \country{France}}

\renewcommand{\shortauthors}{Alvim, Chatzikokolakis, Kawamoto, and Palamidessi}

\begin{abstract}
A common goal in the areas of secure information flow and privacy is to build effective defenses against unwanted leakage of information. 
To this end, one must be able to reason about potential attacks and their interplay with possible defenses. 
In this paper, we propose a game-theoretic framework to formalize strategies of attacker and defender in the context of information leakage, and provide a basis for developing optimal defense methods. 
A novelty of our games is that their utility is given by information leakage, which in some cases may behave in a non-linear way. 
This causes a significant deviation from classic game theory, in which utility functions are linear with respect to players' strategies. 
Hence, a key contribution of this paper is the establishment of the foundations of information leakage games. 
We consider two kinds of games, depending on the notion of leakage considered.
The first kind, the \emph{\qif-games}, is tailored for the theory of quantitative information flow (QIF). 
The second one, the \emph{\edp-games}, corresponds to differential privacy (DP).
\end{abstract}



\maketitle

\section{Introduction}
\label{sec:introduction}
A fundamental problem in computer security is the leakage of sensitive 
information due to the correlation of \emph{secret information} with 
\emph{observable information} 
publicly available, 
or in some way accessible, to the \emph{attacker}. 
Typical examples are \emph{side-channel attacks}, 
in which (observable) physical aspects of the system, such as the execution 
time of a decryption algorithm, may be exploited  by the attacker to
restrict the range of the possible (secret) decryption keys. 
The branch of security that studies the amount of information leaked by a system is called 
\emph{quantitative information flow} (QIF), and it has seen growing 
interest over the past decade.
See, for instance, 
\cite{Clark:05:JLC,Kopf:07:CCS,Smith:09:FOSSACS,Alvim:20:Book}.

Another prominent example is the privacy breach in public databases, where even if personal identifiers such as an individual's name and date of birth are removed, the information stored in the records may be enough to uniquely re-identify a person. The problem of privacy has become more  prominent with the advent of Big Data technologies, and the study of attacks and protection methods is a very active field of research. We mention in particular the \emph{differential privacy} approach \cite{Dwork:06:TCC,Dwork:06:ICALP}, which has become extremely popular in the last decade. 

It has been recognized that randomization 
can be very useful  to obfuscate the link between secrets and observables. 
Examples include various anonymity protocols (for instance, the dining cryptographers \cite{Chaum:88:JC} and Crowds \cite{Reiter:98:TISS}), and  the typical mechanisms for differential privacy such as Laplace and geometric noise \cite{Dwork:06:TCC}.
The \emph{defender} (the system designer or the user) is, therefore, typically probabilistic. 
As for the \emph{attacker} (the party interested in inferring sensitive information),
most works in the literature on quantitative information flow
consider only \emph{passive attacks}, 
limited to observing the system's behavior. 
Notable exceptions are the works of Boreale and Pampaloni \cite{Boreale:15:LMCS}
and of Mardziel et al. \cite{Mardziel:14:SP}, which consider \emph{adaptive attackers} 
who interact with and influence the system.
We note, however, that \cite{Boreale:15:LMCS} does not consider
probabilistic  strategies for the attacker.
As for \cite{Mardziel:14:SP}, although their model allows them, 
none of their extensive  case studies uses probabilistic attack strategies to maximize 
leakage.
This may seem surprising, since, as mentioned before, randomization is known to be 
helpful (and, in general, crucial) for the defender to undermine the attack and protect the secret.
Thus there seems to be an asymmetry between attacker and defender with respect to
probabilistic strategies. 
Our thesis is that there is indeed an asymmetry, but that does not mean that the 
attacker has nothing to gain from randomization: 
when the defender can change his own strategy according to the attacker's 
actions, it becomes advantageous for the attacker to try to be \emph{unpredictable}
and, consequently, adopt a probabilistic strategy. 
For the defender, while randomization is useful for the same reason, it is also 
useful because \emph{it potentially reduces information leakage},
which constitutes the utility of the attacker.
This latter aspect introduces the asymmetry mentioned above. 

In the present work, we consider scenarios in which both attacker 
and defender can make choices that influence the system during an attack. 
We aim, in particular, at analyzing the attacker's strategies that can 
maximize information leakage, and the defender's most appropriate strategies to 
counter the attack and keep the system as secure as possible. 
As argued before, randomization can help both the attacker and defender 
by making their moves unpredictable.  
The most suitable framework for analyzing this kind of interplay is, 
naturally, game theory. In particular, we consider zero-sum games, 
and employ measures of \emph{information leakage} as the \emph{utility} 
function  (with the positive sign for the attacker and the negative one for 
the defender).  
In this context, randomization is naturally captured by
the notion of \emph{mixed strategies}, 
and the optimal strategies for the defender and attacker  
are expressed by the notion of  \emph{Nash equilibria}.

However, it is important to note that 
the notion of
information  is fundamentally different from the typical utility functions modeled in game theory,
such as money, time, or similar resources.
These traditional quantities are linear with respect to the combination of strategies; 
hence the utility of a mixed strategy can be soundly defined as expectation. 
In classical game theory, this concept of utility has been formalized by   the axioms of von Neumann and Morgenstern~\cite{VonNeumann:47:Book}. 
For instance, consider a lottery where the prize is stashed in two identical treasure chests, 
one containing 100 gold coins, and the other containing 200 gold coins, 
and assume that when a participant wins, a hidden fair coin is flipped to decide what chest he should get his prize from.
The winner  expects to get 150 golden coins, and the corresponding benefit from that
does not depend on his exactly knowing
from which chest the prize is coming.
Note that that is still true even if he never
gets to explicitly know the result of the coin flip,
nor of what was the chosen chest: no matter whence the golden coins came, 
the corresponding (monetary) utility to him is the same.

On the other hand, the concept of  information has a rather different nature, which relates to the asymmetry mentioned before.
To understand this point, consider a scenario in which we can pose \qm{yes}/\qm{no} 
questions to one of two oracles and our utility is
the amount of information we can extract from that oracle. 
Assume we know that one of the oracles always tells the truth, 
while the other always lies. 
For instance, if we ask \qm{Is Alice single?} to the truthful oracle and get \qm{yes} as an answer, 
we know for sure that Alice is single, and if we get the same answer from the lying oracle, we know for sure that Alice has a partner.
Note that the  information utility of the answers is maximum, 
as long as we know the oracle they came from, because they
entirely determine the truth or falsehood of the fact we asked about.
Now suppose that we cannot directly query any of these perfectly informative oracles, 
but instead we have to ask our question to some intermediate person who will first flip 
a hidden, fair coin to decide which oracle to pose the question to. 
The person then poses the question to the oracle, hidden from us, obtains an answer, and 
lets us know that the answer was, say, \qm{yes}.
How useful is this answer? 
Not useful at all, indeed, as we don't know what oracle it came from!
The critical insight is that, contrarily to what happens with quantities like money,
time, or other similar resources, the \qm{average} of two highly informative answers is 
not necessarily a highly informative answer.
More precisely, \emph{information is non-linear} with respect to hidden choice (in this example, the hidden coin flip that decides among strategies).
For this reason, traditional game theory as formalized by the von Neumann and Morgenstern's axioms fails to adequately capture games
in which utility is information.

Consequently, in this paper, we consider a new kind of games, which we call 
\emph{information leakage games}.
In these games, the defender can mix two (pure) strategies, by choosing one strategy or the other, without necessarily revealing his choice to the attacker. 
As seen in the above example, the lack of knowledge of the  defender's choice may reduce  the amount of information that the attacker can infer. 
Namely,  the information leaked with a mixed strategy is 
never larger
than the leakage of each original pure strategy,
which means that utility is a \emph{quasi-convex} function of  the defender's strategy. 
As for the attacker's strategies, the properties of the utility depend on the particular definition of leakage, and we will discuss the various instances   in the next section. 
In any case, our setting departs from the standard game theory, in which utility is a linear function on the strategies of both players.
Nevertheless, we show that our games still have Nash equilibria, namely pairs of defender-attacker strategies where neither player has anything to gain by unilaterally changing his own strategy.  
This means that  there exists an ``optimal'' balance of strategies  between defender and attacker, and we will  propose algorithms to compute these  equilibria using a more efficient  variant of the well known subgradient  method  for two variables. 
The efficiency improvement is achieved thanks to the specificity of our problem,  and it can be considered our original contribution. Specifically, we are able to rewrite the minimax problem in 
    a more convenient form (c.f. Proposition~\ref{prop:minproblem}), involving a function which can be locally  maximized easily.
    We then prove that the two standard sufficient conditions  for the convergence of the subgradient method
    (bounded subgradients and bounded distance between the initial and optimal points)
    hold in our case. 
    Finally, we show  that not only our method converges to a saddle point, but also that an approximate solution for the strategy provides an ``approximate saddle point'' (c.f. Theorem~\ref{prop:Lipschitz}).

\subsection{Specific information-leakage frameworks}

The literature  on information leakage in the probabilistic setting can be divided into two main lines of research: 
\emph{quantitative information flow} (QIF)~\cite{Kopf:07:CCS,Smith:09:FOSSACS,Alvim:12:CSF,Alvim:20:Book} and 
\emph{differential privacy} (DP)~\cite{Dwork:06:TCC,Dwork:09:STOC,Dwork:11:CACM}. 
Our game-theoretic approach encompasses both of them, and we will refer to them specifically as \emph{\qif-games} and \emph{\edp-games}, respectively.
For reasoning about information leakage, we adopt the information-theoretic view, which is very popular in the  QIF community. 
In this setting, a system  or a mechanism is modeled as an information-theoretic channel, where the secrets are the inputs and the 
observables are the outputs. We will use this model both for \qif-games and for \edp-games.
We assume that both attacker and defender can influence with their actions the choice of the channel used in the game. 
In other words, we assume that channels are determined by two parameters that represent the actions, or \emph{strategies}, of the two players. 

In the case of QIF, the typical measures of leakage are based on the concept of \emph{vulnerability}, 
which quantifies how easily the secret can be discovered (and exploited) by the attacker. 
There are various models of attackers and corresponding notions of vulnerability   proposed in the literature, but  
for the sake of generality, here we abstract from the specific model. Following \cite{Boreale:15:LMCS,Alvim:16:CSF}, 
we adopt  the view that vulnerability is any convex and continuous function. 
Such a class of functions has been shown in \cite{Alvim:16:CSF} to be the unique family presenting a set of
fundamental information-theoretic properties, and  subsuming most 
previous measures of the QIF literature, 
including \emph{Bayes vulnerability} (a.k.a. min-vulnerability~\cite{Smith:09:FOSSACS,Chatzikokolakis:08:JCS}), 
\emph{Shannon entropy}~\cite{Shannon:48:Bell}, 
\emph{guessing entropy}~\cite{Massey:94:IT}, and 
\emph{$g$-vulnerability}~\cite{Alvim:12:CSF}.
It is important to note that quasi-convexity is implied by  convexity, and therefore this 
definition of vulnerability respects the assumption of our games. 
As for the attacker,  he knows his own choice, thus the information that he can infer by mixing two strategies is the average of that of the individual strategies, 
weighted on the percentages by which they appear in the mix. 
This means that utility is a \emph{linear} (more precisely, \emph{affine}) function of his strategies.  
In conclusion, in the \qif-games the utility is convex on the defender and linear on the attacker.

On the other hand, the property of DP is usually defined for a (probabilistic) mechanism associated with a specific query on datasets. 
We say that a mechanism $\calk$ is $\varepsilon$-differentially private if 
for any two neighbor datasets $x$ and $x'$ that differs only for one record, 
the ratio of the probabilities that $\calk$, applied  respectively on  $x$ and $x'$, reports the same  answer $y$, is bounded by $e^\varepsilon$. 
Here we  adopt a more abstract view: 
Following the approach in \cite{Chatzikokolakis:14:CONCUR,Kifer:14:TDS}, we assume that $x$ and $x'$ range over  a generic domain $\calx$ endowed with a binary relation $\sim$ that generalizes the neighborhood relation. 
For the sake of uniformity, we will also model the  mechanisms as information-theoretic channels, whose typical element is the probability of reporting a certain $y$ when applied to  a given $x$. 
In the differential-privacy literature, the parameter
	$\varepsilon$ is used to control the privacy provided
	by a mechanism: the higher its value, the more secret
	information the  mechanism can reveal.
	Here we follow that approach and adopt as the measure of payoffs the minimum parameter $\varepsilon$ for which a mechanism is $\varepsilon$-differentially private. 

For what concerns the composition of strategies, we remark an important difference between the QIF and DP settings. 
While the measure of leakage in QIF  
can be --and often is-- defined as
an average on all secrets and observables, in DP it is a worst-case measure, in the sense that 
only the worst possible case counts (namely, the minimum $\varepsilon$). 
In line with this philosophy,  the composition of two mechanisms  in DP is not linear with respect to the leakage, even in the case of visible choice: 
the  leakage is determined by the worst of the two, at least for non-trivial combinations. We will call this property \emph{quasi-max} 
and we will see that it is a particular case of quasi-convexity. 
Because of this general non-linearity, it makes sense to explore for DP  also the case in which the defender uses visible choice (i.e., 
he makes known to the attacker which mechanism he is using). In the case of DP, we therefore consider two cases for the defender's choice: visible and invisible. 
For the attacker, on the other hand, we  consider only visible choice, as it is natural to assume that he knows what choices he has made. 
We  show that also for DP, and in all these possible scenarios, Nash equilibria exist, and we  provide algorithms to compute them.

\begin{table}[t]
\centering
{
\small 
\begin{tabular}{c|l|l|}
\cline{2-3}
        &     &      \\[-1.5ex]
       & Defender's choice is visible    & Defender's choice is hidden      \\[0.5ex]
        \hline
\multicolumn{1}{|c|}{\qif-games} & \begin{tabular}[c]{@{}l@{}}\\[-2ex] Utility is linear on attacker and defender.\\ Standard games. See \cite{Alvim:18:Entropy}.\\ \ \\[-2ex] \end{tabular} & \begin{tabular}[c]{@{}l@{}}\\[-2ex]Utility is linear on attacker \\ and convex on defender.\\ \ \\[-1ex] \end{tabular}          \\ \hline
\multicolumn{1}{|c|}{\edp-games}  & \begin{tabular}[c]{@{}l@{}}\\[-2ex] Utility is quasi-max on attacker and defender.\\ \ \\[-2ex]  \end{tabular}    & \begin{tabular}[c]{@{}l@{}}\\[-2ex]Utility is quasi-max on attacker \\ and quasi-convex on defender.\\ \ \\[-2ex] \end{tabular} \\ \hline
\end{tabular}
}
\caption{Various kinds of leakage games. The choice of the attacker is always visible.  }
\label{tab:cases}
\end{table}

Note that we do not consider the case of visible choice for the defender in QIF  because in that setting, for visible choice, the leakage is linear. Since the 
leakage would be linear in both the attacker's and defender's strategies, the games would be  just ordinary games, and the existence of Nash equilibria and 
their computation would derive from standard game theory. The interested reader can find an analysis of this kind of game in \cite{Alvim:18:Entropy}.
Table~\ref{tab:cases} summarizes the various scenarios.

\subsection{Contributions}
The main contributions of this paper are the following: 
\begin{itemize}
\item We define a general framework of \emph{information leakage games}
to reason about the interplay between attacker and defender in 
QIF and DP scenarios. The   novelty w.r.t. standard game theory is that 
the utility of mixed strategies is not 
the expected utility of pure strategies.

\item We extend the existence of  Nash equilibria to \edp-games (their existence for \qif-games is proved in a preliminary version \cite{Alvim:17:GameSec} of this paper). 

\item We provide  methods  for finding Nash equilibria of 
\qif-games by solving a convex optimization problem, and  
for computing optimal strategies for \edp-games 
by solving a sequence of linear programs.
Our implementation of those algorithms for solving \qif-games and \edp-games is freely available online as part of the LIBQIF tool~\cite{libqif}.

\item We examine the difference between our information leakage games and standard game theory, from a principled point of view, 
by analyzing the axioms of von Neumann and Morgenstern that constitute the foundations of the latter, and show why they cannot encompass the notion of utility as information.

\item As a case study on \qif-games, we consider the 
Crowds protocol in  a MANET (Mobile Ad-hoc NETwork).
We study the case in which the attacker can add a 
corrupted node as an attack, the defender can add an honest 
node as a countermeasure, and we compute the defender  component of the  Nash equilibrium.
\item As a case study on \edp-games, we illustrate how to design a local DP mechanism using a real dataset on criminal records. 
Specifically, we consider a \edp-game between a defender who discloses his non-sensitive attributes to a data curator, and an attacker (a data analyst) who queries the data curator about one of the non-sensitive information to extract the defender's secret that is correlated with the disclosed non-sensitive information.
By computing the Nash equilibrium of this \edp-game, we construct a privacy mechanism for the defender.
\end{itemize}

\subsubsection{Relation with preliminary version}

A preliminary version of this work, considering  only the \qif-games, appeared in~\cite{Alvim:17:GameSec}.
The main novelties of this  paper to that version are:
\begin{itemize}
\item We revise and present in more detail the algorithms for computing the Nash equilibria  (Section~\ref{sec:qif-games-computation}). 
\item We introduce the notion of  \edp-games (Section~\ref{sec:dp-games-definition}) and study its theory, showing the existence of Nash equilibria (Sections~\ref{sub:compare:QIF:DP} to~\ref{sec:dp-games-quasi-convexity}).
\item We provide an algorithm for computing optimal strategies for  \edp-games   (Section~\ref{sec:dp-games-computation}). 
\item We show a case study on \edp-games to illustrate how to design a local DP mechanism for an adaptive disclosure of non-sensitive information correlated with a secret (Section~\ref{sub:case-study-dp}).
\item We present proofs for our technical results (Appendix~\ref{sec:proofs}). 
\end{itemize}

\subsubsection{Relation with our subsequent work}
There are two subsequent papers based on the preliminary conference
version~\cite{Alvim:17:GameSec} of this paper.
These were published in~\cite{Alvim:18:POST} (preliminary conference version) 
and in~\cite{Alvim:18:Entropy} (full journal version). 
In these two papers, 
the focus was on the relation between the \qif-games that we consider here and other kinds of games (sequential games, games with visible choice for the defender). In particular, we built a taxonomy of the various kinds of games. 

In the present paper, on the contrary, the focus is on the foundational aspects of a class of games that subsumes QIF and differential privacy (DP) 
and the contrast with the foundations of traditional game theory (as established by the von Neumann-Morgenstern axioms).
Furthermore, the extension to the case of DP  is new to this paper, and was not present in the previous papers \cite{Alvim:17:GameSec}, \cite{Alvim:18:POST}, and \cite{Alvim:18:Entropy}.

\subsection{Related work}
\label{sec:related-work}
There is extensive literature on game theory models for security 
and privacy in computer systems,
including 
network security,
physical security,
cryptography,
anonymity, 
location privacy,
intrusion detection, and 
economics of security and privacy.
See \cite{Manshaei:13:ACMCS} for a survey. 

In many studies, security games have been used to model and analyze utilities 
between interacting agents, especially   attackers and   defenders.
In particular, Korzhyk et al. \cite{Korzhyk:11:JAIR} present 
a theoretical analysis of security games and
investigate the relation between Stackelberg and 
simultaneous games under various forms of uncertainty.
In application to network security,
Venkitasubramaniam~\cite{Venkitasubramaniam:12:ACMTN} investigates 
anonymous wireless networking,     formalized as a  
zero-sum game between the network designer and the attacker.
The task of the attacker is to choose a subset of nodes to monitor 
so that the anonymity of routes is minimum whereas the task of the 
designer is to maximize anonymity by choosing nodes to 
evade flow detection by generating independent transmission schedules.

Khouzani et al.~\cite{Khouzani:15:CSF} present a  
framework for analyzing a trade-off between usability and 
security.
They analyze guessing attacks and derive the optimal policies 
for secret picking as Nash/Stackelberg equilibria.
Khouzani and 
Malacaria \cite{Khouzani:16:CSF,Khouzani:17:CSF,Khouzani:18:Entropy}
study the problem of optimal channel design
in presence of hard and soft constraints. 
They employ entropies (the dual of vulnerability) and, 
in order to capture the widest class of leakages, they consider the property of  core-concavity, a generalization of concavity which 
includes entropies which are not concave (like for instance the R\'enyi entropies when $\alpha>1$).
They show  the existence of universally optimal strategies in the context of a two-player zero-sum game with 
incomplete information and
that the defender's Bayes-Nash equilibrium strategies are solutions of   convex programming problems.  
Furthermore, they show that for any choice of the entropy measure, this problem 
can be solved via convex programming with zero duality gap, for which the 
Karush-Kuhn-Tucker (KKT) conditions can be used.
In summary, their work uses game theory to support the design of optimal channels within operational constraints, while ours 
focuses on modeling the interplay between attacker and defender regarding information leakage of given channels, and to reason about their optimal strategies. 

Concerning costs of security,
Yang et al.~\cite{Yang:12:POST} propose a 
framework to analyze user behavior in anonymity networks. 
The utility is modeled as a combination of weighted cost and anonymity utility. 
They also consider incentives and their impact on users' cooperation.

Some security games have considered leakage of information about the 
defender's choices.
For example, Alon et al.~\cite{Alon:13:SIAMDM} present two-player zero-sum games where a defender chooses probabilities of secrets while an attacker chooses and learns some of the defender's secrets.
Then they show how the leakage on the defender's secrets influences the defender's optimal strategy.
Xu et al.~\cite{Xu:15:IJCAI} present zero-sum security games where the attacker 
acquires partial knowledge on the security resources the defender is protecting, and show 
the defender's optimal strategy under such attacker's knowledge.
More recently, Farhang et al.~\cite{Farhang:16:GameSec} present two-player games 
where utilities are defined taking account of information leakage, although the defender's 
goal is different from our setting.
They consider a model where the attacker incrementally and stealthily obtains 
partial information on a secret, while the defender periodically changes the 
secret after some time to prevent a complete compromise of the system.
In particular,  the defender is not attempting to minimize the leak of a certain secret, but only to make it useless (for the attacker).  
Hence their model of defender and utility  is totally different from ours.

The research area of quantitative information flow (QIF) has been dedicated to the development of theories to quantify and mitigate the amount of information leakage,
including foundational works 
\cite{Clark:05:JLC,Kopf:07:CCS,Smith:09:FOSSACS,Chothia:13:CSF,Boreale:15:MSCS,Americo:20:TIT},
verification of QIF properties \cite{Clark:07:JCS,Kopf:10:CSFa,Chothia:13:CAV,Chothia:14:esorics,Biondi:19:FAOC}, 
and mitigation of information leakage \cite{Khouzani:17:CSF,Khouzani:18:Entropy,Khouzani:19:TIT,Americo:19:CSF}.
In this research area, several studies~\cite{Mardziel:14:SP,Boreale:15:LMCS,Alvim:12:JCS,Kawamoto:15:QAPL,Mestel:19:CSF} have quantitatively modeled and analyzed the information leakage in interactive systems by using different approaches than ours.
Boreale and Pampaloni~\cite{Boreale:15:LMCS} model adaptive attackers interacting with systems and investigate the relationship between adaptive and non-adaptive adversaries, although they do not deal with probabilistic strategies for the attacker.
Mardziel et al.~\cite{Mardziel:14:SP} formalize adaptive attackers using probabilistic finite automata and analyze the information leakage of several case studies using a probabilistic programming language, whereas they do not show any extensive case studies that use probabilistic attack strategies to maximize leakage.
Alvim et al.~\cite{Alvim:12:JCS} show that interactive systems can be modeled as information-theoretic channels with memory and feedback without using game theory to model the adaptiveness of agents.
Kawamoto and Given-Wilson~\cite{Kawamoto:15:QAPL} show a quantitative model of information leakage in scheduler-dependent systems, although they assume only a passive observer who receives traces interleaved by some scheduler.

To the best of our knowledge, however, no other work has explored games with utilities defined as information-leakage measures, 
except the preliminary version of this paper \cite{Alvim:17:GameSec}, which introduced the \qif-games (called information-leakage games there), 
and  our subsequent work~\cite{Alvim:18:POST,Alvim:18:Entropy}, which 
extended the \qif-games  to sequential games, and considered also the case in which the defender's choice is visible to the attacker.

Finally, in game theory, Matsui ~\cite{Matsui:89:GEB} uses the term ``information leakage game'' 
with a meaning different than ours,
namely, as a game in which (part of) the strategy of one player may be leaked in advance to the other player, and the latter may revise his strategy based on this
knowledge.

\subsection{Plan of the paper}

In Section~\ref{sec:preliminaries}, we review fundamental concepts from game theory, quantitative information flow, and differential privacy.
In Section~\ref{sec:qif-games}, we formalize the concept
of \qif-games, in which utility is the vulnerability 
of a channel, and provide a series of examples of such games.
We then  derive the existence of Nash equilibria for these games and provide a method for computing them. 
In Section~\ref{sec:dp-games}, we formalize the concept
of \edp-games, in which utility is the level $\varepsilon$ of differential privacy provided by a system.
We demonstrate that such utility functions are quasi-convex or quasi-max, depending, respectively, on whether
the defender's action is hidden from or visible to the attacker. We derive
the existence of Nash equilibria for these games and provide a method to compute them.
In Section~\ref{sec:comparison-standart-gt}, we compare
our information leakage games with the axiomatic formulation of standard game theory, 
and show why the latter does not  capture our games.
In Section~\ref{sec:case-study}, we apply our framework 
to two case studies: a version of the Crowds protocol and a design of local DP mechanisms for an adaptive disclosure of information correlated with secrets.
Finally, in Section~\ref{sec:conclusion}, we present our final remarks.
The proofs of the formal results are in Appendix~\ref{sec:proofs}.

\section{Preliminaries}
\label{sec:preliminaries}
In this section, we review some basic notions from game theory,
quantitative information flow (QIF), and differential privacy (DP).

We use the following notation.
Given a set $\cali$, we denote 
by $\distr\cali$ the \emph{set of all probability distributions}
over $\cali$.
Given $\mu\in \distr\cali$, its \emph{support}
$\supp{\mu}$ is the set of its elements with positive probabilities, i.e.,
$\supp{\mu} = \{ i \in \cali : \mu(i)>0 \}$.
We write $i \samplefrom \mu$ to indicate that a value 
$i$ is sampled from a distribution~$\mu$.

\subsection{Convexity and quasi-convexity}
We recall that a set $\cals\subseteq\reals^n$ is \emph{convex} if $ps_0 + (1-p)s_1 \in\cals$ for all
$s_0, s_1\in\cals$ and $p\in[0,1]$.
Let   $S$ be a convex set. A function $f:\cals\rightarrow\reals$ is \emph{convex} if $f(ps_0 + (1-p)s_1) \le pf(s_0) + (1-p)f(s_1)$ for all $s_0, s_1\in\cals$ and $p\in[0,1]$,
and it is \emph{concave} if $-f$ is convex. We say that
 $f$ is \emph{quasi-convex} if $f(ps_0 + (1-p)s_1) \le \max\{ f(s_0) , f(s_1) \}$ for all $s_0, s_1\in\cals$ and $p\in[0,1]$, and it is \emph{quasi-concave} if $-f$ is quasi-convex.
 It is easy to see that  if $f$ is  {convex} then it is also   {quasi-convex}, and if it is {concave} then it is also   {quasi-concave}.

\subsection{Two-player, simultaneous games}
\label{subsec:two-player-games}

We recall basic definitions from \emph{two-player games},
a model for reasoning about the behavior of two strategic players.
We refer to \cite{Osborne:94:BOOK} for more details. 

In a game, each player disposes of a set of \emph{actions} 
that he can perform, and obtains some \emph{payoff} (gain or loss)
depending on the outcome of the actions chosen by both players.
The payoff's value to each player is evaluated using a \emph{utility function}.
Each player is assumed to be \emph{rational}, i.e.,  
his choice is driven by the attempt to maximize his own utility.
We also assume that 
the set of possible actions and the utility functions
of both players are \emph{common knowledge}.

In this paper, we only consider \emph{finite games}, namely the cases in which the 
set of actions available to each player is finite.   
Furthermore, we only consider simultaneous games, in which each player chooses actions without knowing the actions chosen by the other.

Formally, a finite simultaneous game between a \emph{defender} and an \emph{attacker}
is defined as a tuple
$(\cald, \cala, \payd, \paya)$, where
$\cald$ is a nonempty set of the \emph{defender's actions}, 
$\cala$ is a nonempty set of the \emph{attacker's actions},
$\payd: \cald\times\cala \rightarrow \reals$ is the 
\emph{defender's utility function}, and
$\paya: \cald\times\cala \rightarrow \reals$ is the 
\emph{attacker's utility function}.

Each player may choose an action deterministically or probabilistically.
A \emph{pure strategy} of the defender (resp. attacker) is a deterministic 
choice of an action, i.e., an element $d\in\cald$ (resp. $a\in\cala$). 
A pair $(d, a)$ of the defender's and attacker's pure strategies is called a \emph{pure strategy profile}.
The values $\payd(d, a)$ and $\paya(d, a)$ 
respectively represent the defender's and the attacker's utilities. 

A \emph{mixed strategy} of the defender (resp. attacker) is a probabilistic choice of an action, 
defined as a probability distribution 
$\delta\in\distr\cald$ (resp. $\alpha\in\distr\cala$).
A pair  $(\delta, \alpha)$ of the defender's and attacker's mixed strategies is called 
a \emph{mixed strategy profile}.
The defender's and the attacker's \emph{expected utility functions} for  mixed strategies are respectively defined as:
\begin{align*}
\Payd(\delta,\alpha) 
\!\eqdef\!{\displaystyle\!\expectDouble{d\leftarrow\delta}{a\leftarrow\alpha}\hspace{-0.5ex}  \payd(d, a)}
=\hspace{-0.5ex} \sum_{\substack{d\in\cald\\ a\in\cala}} \delta(d) \alpha(a) \payd(d, a),
& \hspace{4ex}
\Paya(\delta,\alpha) 
\!\eqdef\!{\displaystyle\!\expectDouble{d\leftarrow\delta}{a\leftarrow\alpha}\hspace{-0.5ex}  \paya(d, a)}
=\hspace{-0.5ex} \sum_{\substack{d\in\cald\\ a\in\cala}} \delta(d) \alpha(a) \paya(d, a).
\end{align*}

A defender's mixed strategy $\delta\in\distr\cald$ is a 
\emph{best response} to an attacker's mixed strategy 
$\alpha\in\distr\cala$ if 
$\Payd(\delta, \alpha) = \max_{\delta'\in\distr\cald}\Payd(\delta', \alpha)$.
Symmetrically, $\alpha\in\distr\cala$ is a \emph{best response} to $\delta\in\distr\cald$ 
if $\Paya(\delta, \alpha) = \max_{\alpha'\in\distr\cala}\Payd(\delta, \alpha')$.
A \emph{mixed strategy Nash equilibrium} is a profile $(\delta^*, \alpha^*)$ such that $\delta^*$ is a best response to $\alpha^*$ and vice versa. 
This means that in a Nash equilibrium, 
no unilateral deviation by any single player provides better 
utility to that player.
If $\delta^*$ and $\alpha^*$ are point distributions  concentrated on some pure strategies $d^*\in\cald$ and $a^*\in\cala$, respectively, then 
$(\delta^*, \alpha^*)$ is a \emph{pure strategy Nash equilibrium}, and is denoted by $(d^*, a^*)$. 
While every finite game has a mixed strategy Nash equilibrium, not all games have a pure strategy Nash equilibrium.

\subsection{Zero-sum games and minimax theorem}
\label{subsec:zero-sum-games}
A game $(\cald,\, \cala,\, \payd, \paya)$ is \emph{zero-sum} if for any $d\in\cald$ and any $a\in\cala$,
$\payd(d, a) = -\paya(d, a)$, i.e., the defender's loss is equivalent to the attacker's gain.
For brevity, in zero-sum games we denote by $u$ the attacker's utility 
function $\paya$, and by $U$ the attacker's expected utility $\Paya$.%
\footnote{Conventionally in game theory  the utility $u$ is  set to 
be that of the first player, but we prefer to look at the utility 
from the point of view of the attacker to be in line with the definition of 
utility as \emph{vulnerability}, which we will define in Section~\ref{subsec:qif}.}
Consequently,  the goal of the defender is to minimize $U$, and the goal of the attacker is 
to maximize it. 

In simultaneous zero-sum games, a Nash equilibrium corresponds to a solution of the \emph{minimax} problem, i.e., a mixed strategy profile $(\delta^*, \alpha^*)$ such that 
$U(\delta^*, \alpha^*)=\min_{\delta} \max_{\alpha} U(\delta, \alpha)$.
(This is equivalent to the \emph{maximin} problem,
which looks for $(\delta^*, \alpha^*)$ such that 
$U(\delta^*, \alpha^*)= \max_{\alpha} \min_{\delta} \allowbreak U(\delta, \alpha)$.)
Von Neumann's minimax theorem ensures that such a solution always exists and is stable.

\begin{restatable}[von Neumann's minimax theorem]{thm}{NeumannMinimax}
\label{theo:vonneumann}
Let $\Delta \subset \reals^m$ and $A \subset \reals^n$ be compact convex sets,
and $\Pay: \Delta\times A\rightarrow\reals$ be a continuous function such that
$\Pay(\delta, \alpha)$ is convex in $\delta\in\Delta$ and concave in $\alpha\in A$.
Then it is the case that
\[\min_{\delta\in\Delta} \max_{\alpha\in A} \Pay(\delta, \alpha) = \max_{\alpha\in A} \min_{\delta\in\Delta} \Pay(\delta, \alpha)~.\]
\end{restatable}
Moreover, under the conditions of Theorem~\ref{theo:vonneumann}, there exists a \emph{saddle point} $(\delta^*, \alpha^*)$ s.t., for all $\delta\in\Delta$ and $\alpha\in A$, 
$ \Pay(\delta^{*}, \alpha) \leq \Pay(\delta^*, \alpha^*) \leq \Pay(\delta, \alpha^*)$.

\subsection{Quantitative information flow (QIF)}
\label{subsec:qif}

Next we recall the standard framework of
quantitative information flow, 
which is used to measure the amount of information 
leakage in a system~\cite{Alvim:20:Book}.
Notably, we define notions of ``secret'', 
an adversary's ``prior knowledge'' about 
the secret (or simply, ``prior''), and a 
``vulnerability measure'' to gauge how well 
the adversary can exploit her knowledge about the secret.
We also define ``channels'', probabilistic mappings 
used to model systems with observable behavior that 
changes  the  adversary's  probabilistic  knowledge,  
making  the secret more vulnerable and hence causing information ``leakage''.
Throughout this paper, we remain faithful to terminology
and a presentation style that is well established in 
the field of quantitative information flow~\cite{Alvim:20:Book}.
  
A \emph{secret} is some piece of sensitive information the 
defender wants to protect, such as a user's password, social 
security number, or current geo-location. 
The attacker often has some partial knowledge 
about the secret values, which is represented as a probability
distribution on secrets called a \emph{prior distribution} 
(or simply a \emph{prior}).
We denote by $\calx$ the set of all possible secrets,
and we typically use $\pi$ to denote a prior that belongs to 
the set $\dist{\calx}$ of distributions over 
$\calx$. 

The \emph{vulnerability} of a secret is a measure of the usefulness 
of the attacker's knowledge about the secret. 
In this paper, we consider a very general notion of vulnerability, 
following~\cite{Alvim:16:CSF}, and define a vulnerability measure $\vf$ 
to be any continuous and convex function of type 
$\dist{\calx} \rightarrow \reals$.
It has been shown in~\cite{Alvim:16:CSF} that these functions coincide
with the set of $g$-vulnerabilities, and are, in a precise sense, 
the most general information measures w.r.t. a set of basic 
axioms.~\footnote{
\label{footnote:convex}
More precisely, if posterior vulnerability 
is defined as the expectation of the vulnerability of posterior
distributions, 
the measure respects the data-processing inequality 
and yields non-negative leakage iff
vulnerability is convex.}

Systems can be modeled as information-theoretic channels.
A 
\emph{channel}
$C : \calx \times \caly \rightarrow \reals$ is a function
in which $\calx$ is a finite set of \emph{input values}, $\caly$ is a finite set 
of \emph{output values}, and $C(x,y)$ represents the conditional 
probability of the channel producing output $y \in \caly$ when 
input $x \in \calx$ is provided. 
Every channel $C$ satisfies $0 \leq C(x,y) \leq 1$ for all 
$x\in\calx$ and $y\in\caly$, and $\sum_{y\in\caly} C(x,y) = 1$ for all $x\in\calx$.

A prior distribution $\pi\in\dist{\calx}$ and a channel $C$ 
with inputs $\calx$ and outputs $\caly$ induce a joint distribution
$p_{X,Y}(x,y) = \pi(x)C({x,y})$ on $\calx \times \caly$,
with marginal 
probabilities $p_{X}(x) = \sum_{y} p_{X,Y}(x,y)$ and 
$p_{Y}(y) = \sum_{x} p_{X,Y}(x,y)$, and conditional probabilities 
$p_{X \mid y}(x{\mid}y) = \nicefrac{p_{X,Y}(x,y)}{p_{Y}(y)}$ if $p_{Y}(y) \neq 0$. 
For a given $y$ (s.t. $p_{Y}(y) \neq 0$), the conditional probability distribution $p_{X \mid y}$ is called the 
\emph{posterior distribution} on $\calx$ given $y$.
When clear from the context, we may omit subscripts on probability distributions, writing, e.g., $p(y)$, $p(x,y)$, and
$p(x \,|\, y)$ for $p_{Y}(y)$, $p_{X,Y}(x,y)$, and
$p_{X | y}(x \,|\, y)$, respectively.

A channel $C$ in which $\calx$ is a set of secret values 
and $\caly$ is a set of observable values produced
by a system can be used to model computations on secrets.
Assume that the attacker has prior knowledge $\pi$ about
the secret value, knows how a channel $C$ works, and
can observe the channel's outputs.
Then the effect of the channel 
is to update the attacker's knowledge from the prior $\pi$ to a 
collection of posteriors $p_{X \mid y}$, each occurring 
with probability $p(y)$.%

Given a vulnerability measure $\vf$, a prior $\pi$, and a channel $C$, the \emph{posterior vulnerability} 
$\postvf{\pi}{C}$ is the vulnerability 
of the secret after the attacker has observed the output of 
$C$.
Formally,
\begin{align}
\label{eq:postvf}
\postvf{\pi}{C}
\eqdef&\, \expect_{y \leftarrow p_{Y}} \priorvf{p_{X \mid y}}.
\end{align}

The \emph{information leakage} of 
a channel $C$ under a prior $\pi$ is a comparison between 
the vulnerability of the secret before the system
was run, called the \emph{prior} vulnerability, and the 
posterior vulnerability of the secret.
The leakage reflects how much the observation of the 
system's outputs increases the utility of the attacker's 
knowledge about the secret. 
It can be defined either 
\emph{additively} (by $\postvf{\pi}{C}-\priorvf{\pi}$), or
\emph{multiplicatively} (by $\nicefrac{\postvf{\pi}{C}}{\priorvf{\pi}}$).
Additive leakage is a measure of an adversary's absolute gain of information, and
	the resulting value retains the same meaning
	as the original vulnerability functions, e.g.,
	the adversary's ``monetary gain'', or ``damage caused''.
	Multiplicative leakage, on the other hand, 
	is a dimensionless scalar representing 
	the adversary's relative gain of information,
	usually interpreted as a percentage.

\subsection{Differential privacy (DP)}
\label{sub:dp}

Next we briefly recall the notion of \emph{differential privacy}~\cite{Dwork:06:TCC,Dwork:06:ICALP}.
Consider a system that takes inputs from a 
set $\calx$, 
endowed with an \emph{adjacency relation} 
$\sim \subseteq \calx \times \calx$ representing which pairs of inputs should be considered (almost) \qm{indistinguishable}. 
We say that $x$ and $x'$ are \emph{adjacent} if $x \sim x'$.
Intuitively, in the discrete case, a system (or mechanism) that takes inputs from $\calx$ 
is differentially-private if the probability of its producing any output given input   $x \in \calx$ is 
roughly the same as the probability of its producing the 
same output given any other input  $x' \in \calx$ adjacent to $x$.
Using the channel notation to represent a system$\slash$mechanism, this intuition is formalized as follows.

\begin{Definition}[$\varepsilon$-differential privacy] \label{def:dp} \rm
Let $\varepsilon \ge 0$ and $\sim \subseteq \calx\times\calx$ be an adjacency relation of secrets.
A channel $C: \calx\times\caly\rightarrow\reals$ provides \emph{$\varepsilon$-differential privacy} if for any pair of values $x$, $x'$ s.t. $x\sim x'$, and for any output value $y\in\caly$,
\[
C(x, y) \leq e^\varepsilon C(x', y)~,
\]
where $C(x, y)$ (resp. $C(x', y)$) is the probability of producing $y$ on input $x$ (resp. $x'$).
\end{Definition}

Note that our definition of differential privacy 
is in line with the influential 
Pufferfish framework~\cite{Kifer:14:TDS} by
Kifer and Machanavajjhala, 
in which a mechanism's (secret) input domain can be arbitrary, and the adjacency relation can be customized to explicitly 
define what is considered to be secret.	
Traditional differential privacy is a particular instantiation 
of this definition in which the (secret) input set consists 
in datasets, and the adjacency relation is defined to reflect 
the usual distinction between the presence or absence 
of any single individual in a dataset.	
It is noteworthy that our abstract definition of differential privacy also encompasses \emph{local differential privacy}, first 
formalized by Kasiviswanathan et al.~\cite{Kasiviswanathan:08:FOCS}.
More precisely, one only needs to define the
set of secrets to be, e.g., values of an attribute for a single individual, and
the adjacency relation to be that in which
any two distinct values are adjacent.

\section{\qif-games}
\label{sec:qif-games}
In this section, we investigate \qif-games, in which the utility function is the information leakage of a channel.
We demonstrate that the  utility function is convex  on mixed strategies, and that there exist Nash equilibria for these games.
We then show how to compute these equilibria.

\subsection{Definition of \qif-games}
\label{sec:qif-games-definition}
We  introduce our games through a couple of examples
to motivate the use of game theory. 
In the first example (the two-millionaires problem) 
the utility happens to behave like in standard game theory, 
but it is just a particular case. In general, there is a fundamental difference, 
and we will illustrate this with our second example (the binary-sum game).

\subsubsection{The two-millionaires problem}

The \emph{two-millionaires problem} was introduced by Yao in \cite{Yao:82:FOCS}. 
In the original formulation, there are two \qm{millionaires}, Alex and Don,
who want to discover who is the richest among them, but neither wants to 
reveal to the other the amount of money they have. 

\begin{figure}
\begin{tabular}{cc}
\begin{minipage}{0.47\textwidth}
\centering
\includegraphics[width=0.7\linewidth]{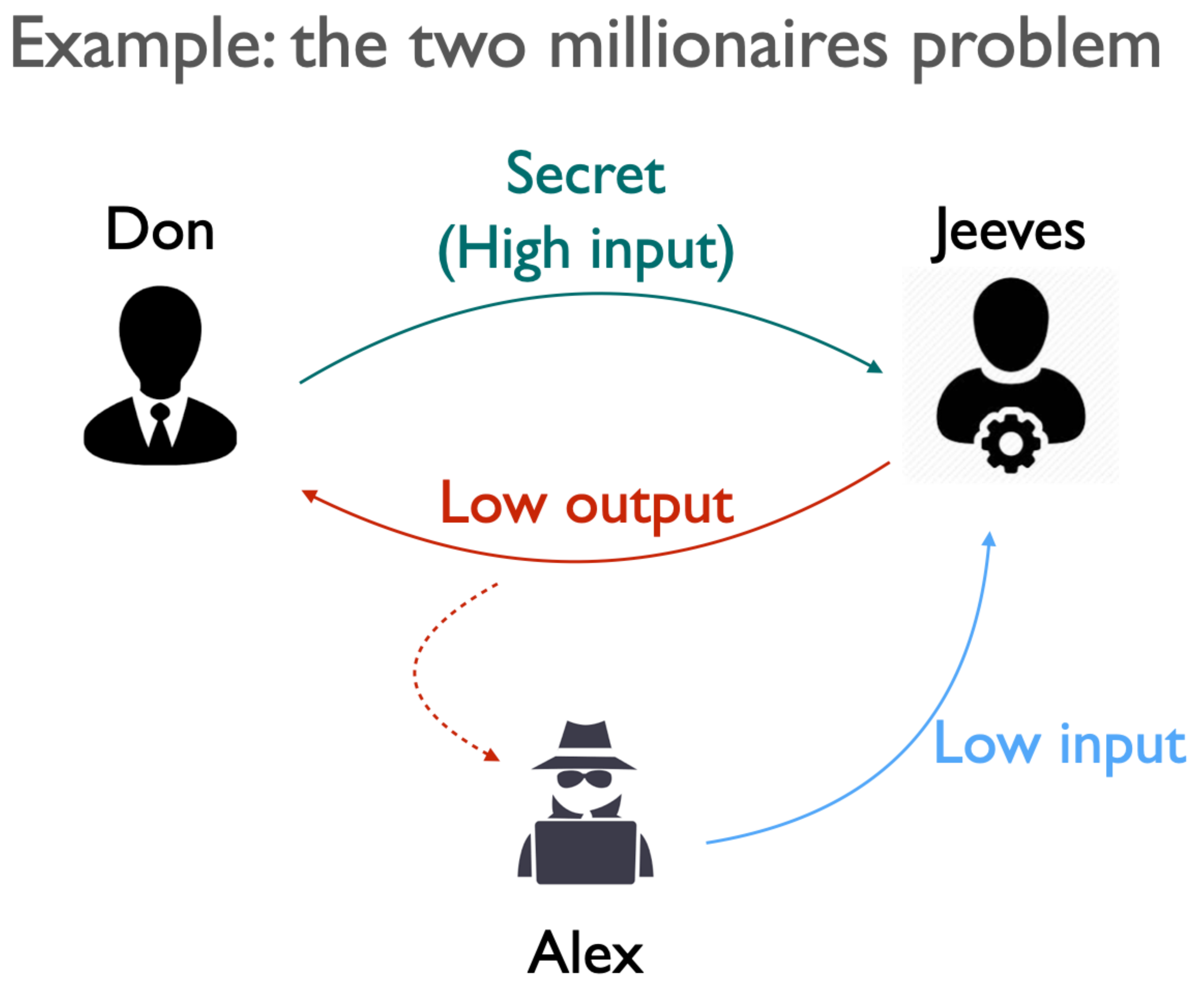}
\caption{Our version of the two-millionaires problem.}\label{fig:millionaires}
\end{minipage}
\hfill
\begin{minipage}{0.51\textwidth}
\centering
\begin{small}
\begin{tabular}{ccc}
\begin{tabular}{l}
\noindent 
\texttt{\underline{Program 0}}\\[2mm]
\noindent \texttt{\textbf{High Input:}} $x \in \{0,1\}$\\
\noindent \texttt{\textbf{Low Input:}} $a \in \{0,1\}$\\
\texttt{\textbf{Output:}} $y\in \{\true,\false\}$\\[1mm]
return $x\leq a$\\[1mm]
\end{tabular}
&
\begin{tabular}{l}
\noindent 
\texttt{\underline{Program 1}}\\[2mm]
\noindent \texttt{\textbf{High Input:}} $x \in \{0,1\}$\\
\noindent \texttt{\textbf{Low Input:}} $a \in \{0,1\}$\\
\texttt{\textbf{Output:}} $y\in \{\true,\false\}$\\[1mm]
return $x\geq a$\\[1mm]
\end{tabular}
\end{tabular}
\end{small}
\caption{The two programs available to Jeeves.}
\label{fig:jeeves}
\end{minipage}
\end{tabular}
\end{figure}

We consider a (conceptually) asymmetric variant of this problem, where Alex 
is the attacker and Don is the defender.
Don wants to learn whether or not he is richer than Alex, 
but does not want Alex to learn anything about the amount $x$ of money he has.
For this purpose, Don sends $x$ to a trusted server, Jeeves, 
who in turn asks Alex, privately, what is her amount $a$ of money.
Jeeves then checks which among $x$ and $a$ is greater, and sends the 
result $y$ back to Don.%
\footnote{The reason to involve Jeeves is that Alex may not want to reveal $a$ to Don, either.}
However, Don is worried that Alex may intercept Jeeves' message containing the 
result of the comparison, and exploit it to learn more accurate information about $x$
by tuning her answer $a$ appropriately (since, given $y$, Alex can deduce whether
$a$ is an upper or lower bound on $x$).
Note that this means, effectively, that $a$ is the choice of action given to Alex in the game.
We assume that Alex may get to know Jeeves' reply, 
but not the messages from Don to Jeeves.\footnote{That may be the case because, say, although all messages are exchanged using encryption, Jeeves' secret key has been hacked 
by Alice, whereas Don's remains protected.}
See Fig.~\ref{fig:millionaires}.

We will use the following information-flow terminology: 
the information that should remain secret (to the attacker) is called \emph{high}, 
and what is visible to (and possibly controllable by) the attacker is called \emph{low}. 
Hence, in the program run by Jeeves $a$ is a \emph{low input} 
and $x$ is a \emph{high input}. 
The result $y$ of the comparison (since it may be intercepted by the attacker) is a \emph{low output}. 
The problem is to avoid the \emph{flow  of information} from $x$ to $y$ (given $a$).

One way to mitigate this problem is to use randomization. 
Assume that Jeeves provides two different programs to ensure the service.
Then, when Don sends his request to Jeeves, he can make a random choice $d$ 
among the two programs 0 and 1, sending $d$ to Jeeves along with the value $x$.
Now, if Alex intercepts the result $y$, it will be less useful to her since
she does not know which of the two programs has been run.
As Don, of course, knows which program was run, the result $y$ will still be 
just as useful to him.~\footnote{Note that $d$ should not be revealed to the attacker:
although $d$ is not sensitive information in itself, knowing it 
would help the attacker figure out the value of $x$.}

In order to determine the best probabilistic strategy that Don should apply to select the program, we  analyze the problem from a game-theoretic perspective. 
For simplicity, we assume that $x$ and $a$ both range in $\{0,1\}$. 
The two alternative programs that Jeeves can run are shown in \Fig{fig:jeeves}.

The combined choices of Alex and Don determine how the system behaves.
Let $\cald = \{0,1\}$ represent Don's possible choices, i.e., the program to run, 
and $\cala = \{0,1\}$ represent Alex's possible choices, i.e., the value of the 
low input $a$. 
We shall refer to the elements of $\cald$ and $\cala$ as \emph{actions}.
For each possible combination of actions 
$d$ and $a$, we can construct a channel 
$C_{da}$
with inputs $\calx = \{0,1\}$ (the set of possible high-input values)
and outputs $\caly = \{\true, \false\}$ (the set of possible 
low-output values),
modeling the behavior of the system \emph{from the point of view of the attacker}.
Intuitively, each channel entry $C_{da}(x,y)$ is the probability that 
the program run by Jeeves (which is determined by $d$) produces output 
$y\in\caly$ given that the high input is $x \in \calx$ and that the 
low input is $a$. 
\Fig{fig:two-millionaries-channels} presents the four 
channel matrices representing all possible behaviors
of the program.
Note that channels $C_{01}$ and $C_{10}$ do not leak any 
information about the input $x$ (output $y$ is constant), 
whereas channels $C_{00}$ and $C_{11}$ completely reveal $x$ (output $y$ 
is in a bijection with $x$).

\begin{figure}[tb]
\begin{center}
{\small
\begin{tabular}{c c c c c}
&
&
$a=0$
&
&
$a=1$
\\[-1ex]
\qquad
\\[-1ex]
$d = 0 \;\; \;(x \leq a?)$
&
\;\;
&
$\begin{array}{|c|c|c|}\hline
C_{00} & y=\true & y=\false \\ \hline
x=0    & 1 & 0 \\
x=1    & 0 & 1 \\ \hline
\end{array}
$
&
\;\;\;\;
&
$\begin{array}{|c|c|c|}
\hline
C_{01} & y=\true & y=\false \\ \hline
x=0    & 1 & 0  \\
x=1    & 1 & 0   \\ \hline
\end{array}
$
\\
\qquad
\\
$d = 1 \;\; \; (x \geq a?)$
&
\;\;
&
$
\begin{array}{|c|c|c|}
\hline
C_{10} & y=\true & y=\false \\ \hline
x=0    & 1 &  0\\
x=1    & 1 & 0 \\ \hline
\end{array}
$
&
\quad
&
$\begin{array}{|c|c|c|}
\hline
C_{11} & y=\true & y=\false \\ \hline
x=0    & 0 & 1 \\
x=1    & 1 & 0 \\ \hline
\end{array}
$
\end{tabular}
}
\end{center}
\caption{The four channels $C_{da}$ representing all possible behaviors of the two-millionaires system, from the point of view of the attacker.}
\label{fig:two-millionaries-channels}
\end{figure}

We now investigate how the defender's and the attacker's strategies
influence the leakage of the system. 
For that, we can consider the (simpler) notion of posterior vulnerability, 
since, for any given prior,
the value of leakage is in a one-to-one, 
monotonic correspondence with the value of posterior vulnerability.~\footnote{More precisely, 
for any fixed prior $\pi$, the additive leakage of a channel
$C$ is obtained by a simple shifting of the 
posterior vulnerability $\postvf{\pi}{C}$ by a term
of $-\priorvf{\pi}$ corresponding to the prior vulnerability.
Similarly, the multiplicative leakage of a channel $C$ 
is obtained by a simple scaling of the posterior
vulnerability $\postvf{\pi}{C}$ by a factor of $\nicefrac{1}{\priorvf{\pi}}$ corresponding to the prior
vulnerability.
Both transformations consist in a monotonic bijection between
the corresponding leakage notions and posterior vulnerability.}
For this example, we consider posterior
Bayes vulnerability~\cite{Chatzikokolakis:08:JCS,Smith:09:FOSSACS}, defined as
\begin{align*}
\postvbayes{\pi}{C} 
\eqdef 
\expect_{y \leftarrow p_{Y}} \max_x p(x \,|\, y)
= \sum_y \max_x \pi(x)C(x,y)~,
\end{align*} 
which
measures the expected probability of the 
attacker guessing the secret correctly in 
one try after having observed the output of the
system.
It can be shown that $\postvbayes{\pi}{C}$ coincides with the converse 
of the Bayes error.

For simplicity, we assume a uniform prior distribution $\pi_u$.
It has been shown that, in this case, the posterior Bayes vulnerability of
a channel $C$ can be computed as the sum of the greatest elements 
of each column of $C$, divided by the size
of the high-input domain~\cite{Braun:09:MFPS}. 
Namely,
\begin{align*}
\postvbayes{\pi_u}{C} = \frac{\sum_{y}\max_xC(x,y)}{|\calx|}~.
\end{align*}
Simple calculations yield the results $\postvbayes{\pi_{u}}{C_{00}}= \postvbayes{\pi_{u}}{C_{11}} = 1$ and $\postvbayes{\pi_{u}}{C_{01}}= \postvbayes{\pi_{u}}{C_{10}} = \nicefrac{1}{2}$,
which are summarized in the utility table 
of \Fig{fig:vulnerabilitygame}, 
which is similar to that of the well-known \qm{matching-pennies} game.

\begin{figure}[tb]
\centering
\renewcommand{\arraystretch}{1.1}
{
\small
\begin{tabular}{c c c c}
The attacker's utility
&
&
The defender's utility
\\[0.3ex]
$
\begin{array}{|c|c|c|}
\hline
\vbayes & a=0 & a=1 \\ \hline
d=0    & 1 & \nicefrac{1}{2} \\ 
\hline
d=1    & \nicefrac{1}{2} & 1\\ \hline
\end{array}
$
&&
$
\begin{array}{|c|c|c|}
\hline
-\vbayes & a=0 & a=1 \\ \hline
d=0    & -1 & -\nicefrac{1}{2} \\ 
\hline
d=1    & -\nicefrac{1}{2} & -1\\ \hline
\end{array}
$
\end{tabular}
}
\renewcommand{\arraystretch}{1}
\caption{Utility tables for the two-millionaires game.
The attacker's and defender's utility values sum to zeros.}
\label{fig:vulnerabilitygame}
\end{figure}

As in standard game theory, there may not exist an optimal pure strategy profile.
The defender, as well as the attacker, can then try to minimize/maximize the system's vulnerability by
adopting mixed strategies $\delta$ and $\alpha$, respectively. A
crucial task is  \emph{evaluating the vulnerability} of the system under such
mixed strategies.
This evaluation is naturally performed from the attacker's point of view,
who  knows his own choice $a$, but \emph{not the defender's choice
$d$} (hidden choice).

To define precisely the vulnerability of a mixed strategy, let us introduce some 
notation: given a channel matrix $C$ and 
a scalar $k$, $k\,C$ is the matrix obtained by multiplying every element of $C$ by $k$.
Given two \emph{compatible} channel matrices 
$C_1$ and $C_2$, namely matrices with the same
indices of rows and columns,\footnote{Note that two channel matrices with different column indices can always be made compatible by adding  appropriate columns with $0$-valued cells in each of them.} $C_1+C_2$ is obtained by adding  the cells of 
$C_1$ and $C_2$ with same indices. 
Note that if $\mu$ is a probability distribution on $\cali$, then 
$\displaystyle{\expect_{i\leftarrow\mu}} C_{i}$ 
is a channel matrix.

From the attacker's point of view, the resulting system is  the convex combination 
\begin{align}
\label{def:hidden-choice}
C_{\delta a} \eqdef \expect_{d\leftarrow\delta} C_{da}\,,
\end{align} 
i.e., a probabilistic choice between
the channels representing the defender's actions.
	
The overall
vulnerability of the system is then given by the vulnerability of $C_{\delta
a}$, averaged over all attacker's actions. 

We now define formally the ideas illustrated above
using a simultaneous game (Section~\ref{subsec:two-player-games}).

\begin{Definition}[\qif-game]\rm
\label{eq:v-mixed}
A \emph{\qif-game} is a 
simultaneous game
$(\cald, \cala, C, \vf)$ where 
$\cald, \cala$ are the finite sets of actions of the attacker and the defender, respectively,  $C=\{C_{da}\}_{da}$ is a family of 
channel matrices of same type indexed on pairs of actions  $d\in\cald,a\in\cala$, and $\vf$ is a vulnerability measure.
The game is zero-sum, and for a prior $\pi$, the utility for 
the attacker of a  pure strategy profile $(d,a)$ is given by $\postvf{\pi}{C_{d a}}$. 
The utility $\vf(\delta,\alpha)$ for the attacker of a 
mixed strategy profile $(\delta,\alpha)$ is  defined as
\[
	\vf(\delta,\alpha)\eqdef \expectDouble{a\leftarrow\alpha}{}\hspace{-0.5ex} \postvf{\pi}{C_{\delta a}} =\smallersum{a} \alpha(a) \,\postvf{\pi}{C_{\delta a}},
\]
where $C_{\delta a}$ is the hidden-choice channel
composition defined as in~\eqref{def:hidden-choice}.
\end{Definition}

In our example, $\delta$ can be represented by a single number $p$: the
probability that the defender chooses $d = 0$ (i.e., Program $0$).
From
the point of view of the attacker, once he has chosen $a$, the system will look
like a channel $C_{pa} = p\, C_{0a} + (1-p)\,C_{1a} $. For instance, in the case $a=0$, if
$x$ is $0$ Jeeves will send $\true$ with probability $1$, but, if $x$ is $1$,
Jeeves will send $\false$ with probability $p$ and $\true$  with probability
$1-p$. Similarly for $a=1$.
\Fig{fig:mixed-two-millionaries-channels} summarizes the various channels
modeling the attacker's point of view.
It is easy to see that $\postvbayes{\pi_{u}}{C_{p0}} = \nicefrac{(1+p)}{2}$ and $\postvbayes{\pi_{u}}{C_{p1}} = \nicefrac{(2-p)}{2}$. In this case $\postvbayes{\pi_{u}}{C_{pa}}$ coincides 
with the expected utility with respect to $p$, i.e., $\postvbayes{\pi_{u}}{C_{pa}} = p\,\postvbayes{\pi_{u}}{C_{0a}}+(1- p)\,\postvbayes{\pi_{u}}{C_{1a}}$. 
 
\begin{figure}[t]
{\small
\centering
$
\begin{array}{|c|c|c|}
\multicolumn{3}{c}{\text{Channel for $a=0$}} \\[1mm]
\hline
C_{p0} & y=\true & y=\false \\ \hline
x=0    & 1 & 0 \\
x=1    & 1-p & p \\ \hline
\end{array}
$
\qquad \qquad
$
\begin{array}{|c|c|c|}
\multicolumn{3}{c}{\text{Channel for $a=1$}} \\[1mm]
\hline
C_{p1} & y=\true & y=\false \\ \hline
x=0    & p & 1-p  \\
x=1    & 1 & 0   \\ \hline
\end{array}
$
}
\caption{Channels representing two-millionaires mixed strategy of the defender, 
	from the point of view of the attacker, where $p$ is the probability the 
	defender picks action $d=0$.}
\label{fig:mixed-two-millionaries-channels}
\end{figure}

Assume now that the attacker chooses $a=0$ with probability $q$ and $a=1$ with probability $1-q$. The utility is obtained as the expectation with respect to the strategy of the attacker, hence the total utility  is:
$\vbayes (p,q) = \nicefrac{q  \,(1+p)}{2} + \nicefrac{(1-q)\,(2-p)}{2}$, which is affine in both $p$ and $q$. By applying standard game-theoretic techniques, 
we  derive that the optimal strategy is $(p^*,q^*) = (\nicefrac{1}{2},\nicefrac{1}{2})$. 

In the above example,  things work just like in standard game theory. 
However, the next example fully exposes the difference of our games with respect to those of  standard game theory.

\subsubsection{The binary-sum game}

The previous example is an instance of a general scenario in which a user, Don, delegates to a server, Jeeves, a certain computation that also requires some input from other users. 
Here we will consider another instance, in which the function to be computed is 
the binary sum $\oplus$.
We assume Jeeves provides the programs in \Fig{fig:oplus}.
The resulting channel matrices are represented in \Fig{fig:oplus-channels}.

\begin{figure}[tb]
\centering
\begin{small}
\begin{tabular}{ccc}
\begin{tabular}{l}
\noindent 
\texttt{\underline{Program 0}}\\[1mm]
\noindent \texttt{\textbf{High Input:}} $x \in \{0,1\}$\\
\noindent \texttt{\textbf{Low Input:}} $a \in \{0,1\}$\\
\texttt{\textbf{Output:}} $y\in \{0,1\}$\\ 
return $x\oplus a$\\[1mm]	
\end{tabular}
&
\qquad \qquad \qquad
&
\begin{tabular}{l}
\noindent 
\texttt{\underline{Program 1}}\\[1mm]
\noindent \texttt{\textbf{High Input:}} $x \in \{0,1\}$\\
\noindent \texttt{\textbf{Low Input:}} $a \in \{0,1\}$\\
\texttt{\textbf{Output:}} $y\in \{0,1\}$\\
return $x\oplus a \oplus1$\\[1mm]
\end{tabular}
\end{tabular}
\end{small}	
\caption{The programs for computing binary-sum, and for computing its complement.}
\label{fig:oplus}
\end{figure}

\begin{figure}[tb]
\begin{center}
{\small 
\begin{tabular}{r c c c c}
&
&
$a=0$
&
&
$a=1$
\\[-1ex]
\qquad
\\[-1ex]
$d = 0 \;\; \;(x \oplus a)$
&
\;\;
&
$\begin{array}{|c|c|c|}\hline
C_{00} & y=0 & y=1 \\ \hline
x=0    & 1 & 0 \\
x=1    & 0 & 1 \\ \hline
\end{array}
$
&
\;\;\;\;
&
$\begin{array}{|c|c|c|}
\hline
C_{01} & y=0 & y=1 \\ \hline
x=0    & 0 & 1  \\
x=1    & 1 & 0   \\ \hline
\end{array}
$
\\
\qquad
\\
$d = 1 \;\; \; (x \oplus a \oplus 1)$
&
\;\;
&
$
\begin{array}{|c|c|c|}
\hline
C_{10} & y=0 & y=1 \\ \hline
x=0    & 0 & 1\\
x=1    & 1 & 0 \\ \hline
\end{array}
$
&
\quad
&
$\begin{array}{|c|c|c|}
\hline
C_{11} & y=0 & y=1 \\ \hline
x=0    & 1 & 0 \\
x=1    & 0 & 1 \\ \hline
\end{array}
$
\end{tabular}
}
\end{center}
\caption{The four channels $C_{da}$ representing all possible
	behaviors of the binary-sum system, from the attacker's point of view.}
\label{fig:oplus-channels}
\end{figure}

We consider again Bayes posterior vulnerability as utility. Simple calculations yield values $\postvbayes{\pi_{u}}{C_{00}}= \postvbayes{\pi_{u}}{C_{11}} = \postvbayes{\pi_{u}}{C_{01}}= \postvbayes{\pi_{u}}{C_{10}} = 1$. 
Thus  for the pure strategies we obtain the  utility table shown in \Fig{fig:vulnerabilitygameoplus}. 
This means that all pure strategies have the same utility $1$ and therefore they are all equivalent.
In standard  game theory 
this would mean that also 
the mixed  strategies have
 the same utility $1$, since they are defined as expectation. 
In our case, however, the utility of a mixed strategy of the defender is convex on the
distribution,~\footnote{The convexity of posterior vulnerability w.r.t. the strategy of the
defender is formally shown in Theorem~\ref{theo:convex-V-q} ahead.} so it may be convenient
for the defender to adopt a mixed strategy. 
Let $p$ and $1-p$ be the probabilities of the defender choosing Program $0$ and Program $1$, respectively. 
From the point of view of the attacker, for each of his choices of $a$,   the system will appear as the probabilistic channel $C_{pa}$ represented in \Fig{fig:mixed-binary-sum}.
\begin{figure}[tb]
\begin{tabular}{cc}
\begin{minipage}{0.46\textwidth}
\centering
{\small 
	\centering
	\renewcommand{\arraystretch}{1.1}
	\[
	\begin{array}{|c|c|c|}
	\hline
	\vbayes & a=0 & a=1 \\ \hline
	d=0    & 1 & 1 \\ 
	\hline
	d=1    & 1 & 1\\ \hline
	\end{array}
	\]
	}
	\renewcommand{\arraystretch}{1}
	\caption{Utility table for the binary-sum game, from the attacker's point of view. The defender's utility values are $-1$ for all strategy profiles; the attacker's and defender's utility values sum to zeros.
	\label{fig:vulnerabilitygameoplus}}
\end{minipage}
\hfill
\begin{minipage}{0.01\textwidth}
~
\end{minipage}
\hfill
\begin{minipage}{0.50\textwidth}
\vspace{-0.8ex}%
\begin{center}
{\small 
\begin{tabular}{ c c}
$a=0$
&
$a=1$
\\[-1ex]
\qquad
\\[-1ex]
$\begin{array}{|c|c|c|}\hline
C_{p0} & y=\true & y=\false \\ \hline
x=0    & p & 1-p \\
x=1    & 1-p & p \\ \hline
\end{array}
$
\;
&
$\begin{array}{|c|c|c|}
\hline
C_{p1} & y=\true & y=\false \\ \hline
x=0    & 1-p & p  \\
x=1    & p & 1-p   \\ \hline
\end{array}
$
\end{tabular}
}
\end{center}
\caption{Channels induced by the binary-sum mixed strategy of the defender, from the point of view of the attacker, where $p$ is the probability the defender picks action $d=0$.}
\label{fig:mixed-binary-sum}
\end{minipage}
\end{tabular}
\end{figure}
Simple calculations yield
$\postvbayes{\pi_{u}}{C_{p0}} = \postvbayes{\pi_{u}}{C_{p1}} = 1-p$ if $p \leq \nicefrac{1}{2}$,
and $\postvbayes{\pi_{u}}{C_{p0}} = \postvbayes{\pi_{u}}{C_{p1}} = p$ if $p \geq \nicefrac{1}{2}$.
On the other hand, with respect to a mixed strategy  of the attacker the utility is still defined as expectation. Since in this case the utility is the same for $a=0$ and $a=1$,   it  remains the same for any strategy of the attacker. Formally, $\vbayes(p,q) = q\, \postvbayes{\pi_{u}}{C_{p0}} + (1-q) \, \postvbayes{\pi_{u}}{C_{p1}} = \postvbayes{\pi_{u}}{C_{p0}}$, which does not depend on $q$ and it is minimum for  $p=\nicefrac{1}{2}$. We conclude that the point of equilibrium is $(p^*,q^*)=(\nicefrac{1}{2},q^*)$ for any value of $ q^*$.

\subsection{Existence of Nash equilibrium for \qif-games}
\label{sec:qif-games-convexity}
In \cite{Alvim:18:Entropy} it was proved that the posterior
vulnerability of a convex combination of  channels  
is smaller than or equal to the  convex combination of their vulnerabilities. 
As a consequence, posterior vulnerability  is a convex function of the strategy of the defender, and using the von Neumann's minimax
theorem we are able to derive the existence of the Nash equilibrium for \qif-games.

\begin{restatable}[Convexity of vulnerability w.r.t.  channel composition \cite{Alvim:18:Entropy}]{thm}{ConvexityChannelCompo}
\label{theo:convex-V-q}
Let $\{C_{i}\}_{i \in \cali}$ be a family of channels of the same type, all with domain $\calx$. 
Then, for every prior distribution $\pi \in \dist{\calx}$, every
vulnerability $\vf$, and
any probability distribution $\mu$ on $\cali$, we have that
\begin{align*}
\vf\bigl[\pi, \expect_{i\leftarrow\mu} C_{i} \bigr] \leq \expect_{i\leftarrow\mu} \postvf{\pi}{C_{i}} {.}
\end{align*}
\end{restatable}

As formalized in the next result, the existence of Nash equilibria for \qif-games immediately follows from the above 
theorem.

\begin{restatable}[Existence of Nash-equilibrium for \qif-games]{cor}{Nash}
\label{cor:Nash}
For any (zero-sum) \qif-game there exists a Nash equilibrium, which in general is given by a mixed strategy.
\end{restatable}

\subsection{Computing equilibria for \qif-games}
\label{sec:qif-games-computation}
Recall that the utility of a mixed strategy profile $(\delta,\alpha)$, given by Definition~\ref{eq:v-mixed}, is
\[
\vf(\delta,\alpha) = \expect_{a\leftarrow\alpha} 
\vf\bigl[\pi, \expect_{d\leftarrow\delta} C_{da} \bigr]\,,
\]
and that $\vf(\delta,\alpha)$ is convex on $\delta$ and affine on $\alpha$. 
Theorem~\ref{theo:vonneumann} guarantees the existence of an equilibrium
(i.e., a saddle-point) $(\delta^*, \alpha^*)$, which is a solution of both the
minimax and the maximin problems. The goal in this section is to compute:
(i) a
$\delta^*$ that is part of an equilibrium, which is important in order to
optimize the defense, and 
(ii) the utility $\vf(\delta^*,\alpha^*)$, which is
important to provide an upper bound on the effectiveness of an attack when
$\delta^*$ is applied.

This is a convex-concave optimization problem for which various methods have been
proposed in the literature. If $\vf$ is twice differentiable (and satisfies a
few extra conditions) then the Newton method can be applied \cite{Boyd:04:BOOK};
however, many $\vf$'s of interest, and notably the Bayes vulnerability, one of the most popular   
vulnerability measures, are not differentiable. For non-differentiable
functions, \cite{Nedic:09:JOTA} proposes an approximate iterative method that 
computes a subgradient and updates  both $\alpha$ and $\delta$ at each step in the direction of the subgradient, 
moving towards the  point of equilibrium. The process stops when the change in the value of 
$\vf(\delta,\alpha)$ induced by the update of $\alpha$ and $\delta$   is ``small enough'', meaning that we are close to $\vf(\delta^*,\alpha^*)$,
and the current values of $\alpha$ and $\delta$ are returned as approximations of 
$\delta^*$ and $\alpha^*$.
Some precautions must be taken to ensure stability, i.e., to avoid  swinging  
around the point of equilibrium: the size of the updates needs to become smaller as we get closer to it.
Clearly, there is a trade-off between the precision of the approximation and the convergence speed. 

In case we are interested in computing only the optimal strategy of the defender, 
however, we propose a method more efficient than the one of \cite{Nedic:09:JOTA}. 
Our method is iterative like the one in  \cite{Nedic:09:JOTA}, but, in contrast to \cite{Nedic:09:JOTA},
 at every step it performs the update only on the $\delta$ component, while 
it computes the exact $\argmax$ on the $\alpha$ component, thus improving efficiency. 
For this purpose, we exploit the fact that 
$\vf(\delta,\alpha)$ is affine on $\alpha$ (not just concave). This means that,  for a fixed $\delta$, maximizing
$\expect_{a\leftarrow\alpha} \postvf{\pi}{\expect_{d\leftarrow\delta} C_{da}}$ 
simply involves
selecting an action $a$ with the highest 
$\postvf{\pi}{\expect_{d\leftarrow\delta} C_{da}}$ and
assigning probability $1$ to it.

\begin{restatable}{propo}{Minproblem}\label{prop:minproblem}
	The minimax problem 
	$
	\delta^* = \argmin_\delta
	\max_\alpha \vf(\delta,\alpha)
	$
	is equivalent to:
	\begin{equation}\label{eq:effe}
	\delta^* = \argmin_\delta f(\delta) \quad \mbox{where} \quad 
	f(\delta) = \max_a\postvf{\pi}{\textstyle\expect_{d\leftarrow\delta} C_{da}}.
	\end{equation}
\end{restatable}
It is important to point out that solving  \eqref{eq:effe} produces a  $\delta^*$ that is 
the defender's mixed strategy of
an equilibrium, but in general it does not produce 
the attacker's mixed strategy $\alpha^*$ of the equilibrium.
For the latter,  we would need to solve the maximin problem.

Note that $f$ is a function of a single variable $\delta$, and our problem reduces to computing
its minimum value.
We now proceed to apply the subgradient descent method for this task.
We start by recalling the notions of subgradient and projection on the simplex. 
\begin{Definition}[Subgradient]\rm
Let $U$ be a convex open set in the Euclidean space  $\mathbb{R}^n$.  A \emph{subgradient} of a convex function $f:U\to \mathbb{R}$ at a point $x_0$ in $U$ is any vector $v\in \mathbb{R}^n$ such that,
for all $x \in U$,
\[ f(x)-f(x_{0})\geq v\cdot(x-x_{0}) \, ,   \]
where ``$\cdot$'' denotes the dot product between vectors. 
\end{Definition}

\begin{Definition}[Projection]\rm
The \emph{$n$-simplex} $\mathbb{S}^n$ is the set of all vectors in $\mathbb{R}^n$ that represent probability distributions. 
Given a vector $v\in \mathbb{R}^n$, the \emph{projection} of $v$ on $\mathbb{S}^n$, which we will denote by $P(v)$, is the probability 
distribution $\delta\in\mathbb{S}^n$ that is closest, in the Euclidean sense, to $v$. 
 \end{Definition}
Note that  if $v\in\mathbb{S}^n$ then $P(v)=v$.
To compute the projection we can use the  efficient algorithms  proposed in
\cite{Wang:2013:arXiv}. 

We are  now ready to define our iterative method for the approximation of  $\delta^*$ and  $\vf(\delta^*,\alpha^*)$. 
The idea is to start from an arbitrary fully supported distribution over $\cald$, for instance the uniform distribution $u_\cald$, and then compute the next $\delta$ 
from previous one following~\eqref{eq:effe}.
More precisely, at step $k+1$ we compute $\delta^{(k+1)}$ from $\delta^{(k)}$ according to the following inductive definition:
\begin{equation}\label{eq:rec}
	\begin{array}{lcl}
	\delta^{(1)} = u_\cald,\\[1ex]
	\delta^{(k+1)} = P(\delta^{(k)} - s_k h^{(k)})\\[1ex]
	\mbox{where } s_k = \nicefrac{0.1}{\sqrt{k}}~
	\mbox{ and $h^{(k)}$ is  any subgradient of $ f(\delta) = \max_a\postvf{\pi}{\expect_{d\leftarrow\delta} C_{da}}$ at point } \delta^{(k)}.\hspace{-6ex} 
	\end{array}
\end{equation}

The \emph{step size}  $s_k$ has to be decreasing in $k$ in order to avoid the swinging effect, as explained above.  
There are  various  choices for  $s_k$  that guarantee convergence \cite{Boyd:06:misc}. 
We have chosen $s_k=\nicefrac{0.1}{\sqrt{k}}$ because we have determined heuristically that it gives a good performance.  

We use the same  stopping criterion of \cite[Section~3.4]{Boyd:06:misc}. Namely, at each step $k$,
we compute the following value $l^{(k)}$, which is guaranteed to be a lower bound on $f(\delta^*)$:
\begin{equation}
	l^{(k)} =
		\frac{ 2\sum_{i=i}^{k} s_i f(\delta^{(i)}) - \frac{|\calx|-1}{|\calx|} - \sum_{i=i}^{k} s_i^2\|h^{(i)}\|_2^2  }
			 {2\sum_{i=i}^{k} s_i }
	~.
\end{equation}
Hence, we can stop the process when 
$f(\delta^{(k)}) - l^{(k)} \le \epsilon$, for some given  $\epsilon>0$, 
and then the value returned is   $\hat\delta = \delta^{(k)}$.\footnote{
	Note that $f(\delta^{k})$ is not necessarily decreasing,
	so an optimization would be to keep track of the \emph{best} values
	of $f(\delta^{(k)})$ and $l^{(k)}$, stop when
	$f(\delta^{(\text{best})}) - l^{(\text{best})} < \epsilon$,
	and return $\hat\delta = \delta^{(\text{best})}$.
}
The parameter $\epsilon$ determines the desired level of approximation,
as  formally stated in Theorem~\ref{prop:Lipschitz} below. Before stating the result we need to recall the Lipschitz  property.

\begin{Definition}[Lipschitz]\rm
Let $(A,d_A)$ and $(B,d_B)$ be two metric spaces and  let $G$ be a non-negative real constant. 
A function $F: A\rightarrow B$ is \emph{$G$-Lipschitz} if 
for all $a,a'\in A$,
\[
d_B (F(a),F(a'))  \le G\cdot d_A(a,a')~.
\]
We also say that $F$ is \emph{Lipschitz} if it is $G$-Lipschitz for some $G$.
\end{Definition}

\begin{restatable}[Convergence]{thm}{Lipschitz}
\label{prop:Lipschitz}
	If  \,$\vf$ is Lipschitz	then the sequence $\{\delta^{(k)}\}_k$ in ~\eqref{eq:rec}
	converges to
	a $\delta^*$ that is part of an equilibrium.
	Moreover, when the stopping condition is met, i.e., $f(\delta^{(k)}) - l^{(k)} < \epsilon$, then
	the approximate solution $\hat\delta =\delta^{(k)}$ returned by the algorithm, and for an equilibrium $(\delta^*, \alpha^*)$,  
	\[
		\vf(\hat\delta, \alpha) - \epsilon
		\le
		\vf(\hat\delta,\alpha^*)
		\le
		\vf(\delta,\alpha^*) + \epsilon
		\qquad \forall \delta,\alpha~.
	\]
	This implies that the computed strategy is $\epsilon$-close to the optimal one, namely:
	\[
		|\vf(\hat\delta,\alpha^*) - \vf(\delta^*,\alpha^*) | \le \epsilon  ~.
	\]
\end{restatable}
 
The convergence crucially depends on the $G$-Lipschitzness assumption, which ensures the boundness of the subgradients and therefore   that there is no “swinging effect” around  the  point  of  equilibrium.
 One may be tempted to also find a relation between the constant $G$  and the convergence rate, but,  
 unfortunately, there is no such relation in general. Essentially, this is because  $G$ is a global bound, while the convergence rate depends on the steepness around the point of equilibrium  and how well it combines with the rate by which the step size diminishes.

Notably, the  Lipschitz condition is satisfied by the large class of the $g$-vulnerability measures~\cite{Alvim:12:CSF} when the ``set of guesses'' (or ``actions'') is finite. 
The set of guesses $\calw$ is a parameter of a $g$-vulnerability measure, along with the gain function $g:\calw\times\calx\rightarrow \reals$. 
One example of $g$-vulnerability   with finite $\calw$ is that of Bayes vulnerability: in this case $\calw=\calx$ and $g(w,x) = 1$ if $w=x$, and $g(w,x) = 0$ otherwise. 

In general, if $\calw$ is finite, the
 posterior $g$-vulnerability $\vf$ is given by:
\[\textstyle
\postvf{\pi}{C} = \sum_{y}\, \max_{w}\, \sum_{x}\, \pi(x) \,C(x,y) \,g(w,x)~.
\]
The above $\vf$ is piecewise linear, and therefore  $G$-Lipschitz, where  $G$ is the maximum of the norms of the subgradients at the point distributions. 
In general, a subgradient vector $h^{(k)}$ is given by:
\[
h^{(k)}_d = \delta^{(k)}(d) \,{\textstyle\sum_y}\,  \pi(x^*_y)\, C_{da^*}(x^*_y,y)~,
\]
where $a^*,x^*_y$ are (any of) the ones giving the max in the branches of
$f(\delta^{(k)})$.

Note  that if $\vf$ is piecewise linear then also $f$ is piecewise linear. Hence, in the case of  $g$-vulnerability,   the convex optimization problem
could be transformed into a linear one using a standard technique, and then solved
by linear programming. 
However, typically $\vf$ has a large number of max branches,
and consequently  this conversion can generate a  huge number
of constraints.
In our experiments, we found that the subgradient method
described above is significantly more efficient than linear programming
(although the latter has the advantage of giving an exact solution).

On the other hand, not all vulnerabilities are Lipschitz; Shannon entropy, for instance,  is not.
Note that Shannon entropy can be expressed as $g$-vulnerability, but the corresponding set of guesses  $\calw$ is infinite~\cite{Alvim:16:CSF}.

Finally, although the subgradient method is general, it might
be impractical in applications where the number of attacker or defender actions
is very large. Application-specific methods could offer better scalability in
such cases; we leave the development of such methods as future work.

\section{\edp-games}
\label{sec:dp-games}

In this section, we investigate \edp-games, in which the utility function is the  level of  privacy 
of a channel, namely the minimum $\varepsilon$ for which the channel is $\varepsilon$-differentially private.

We demonstrate that such utility functions are quasi-convex or quasi-max (rather than linear) on mixed strategies, depending, respectively, on whether
the defender's action is hidden from or visible to the attacker, and derive
the existence of Nash equilibria for these games.
We then show how to compute  these  equilibria.

\subsection{Definition of \edp-games}
\label{sec:dp-games-definition}

We begin by formalizing the level of differential privacy of a channel, following an alternative characterization of differential privacy based on max-divergence~\cite{Dwork:14:Algorithmic}.
Let $\calx,\caly$ be two sets (the domains of \qm{secrets} and \qm{observables}, respectively), and let 
  $\sim$ be a symmetric  binary relation on $\calx$, i.e., $\sim\subseteq \calx \times \calx$ and such that $x\sim x'$  iff $x'\sim x$. 
We say that a channel $C:\calx \times \caly \rightarrow \reals$ is \emph{conforming} to  $\sim$  if 
for all $x$, $x' \in \calx$ and $y \in \caly$, 
$x\sim x'$ implies that $C(x,y) = 0$ iff $C(x',y) = 0$.

Note that the notions of domain and 
adjacency relation are more abstract than in the typical definition of differential privacy. 
This is because we want to  capture both the cases of the standard (central) DP and the local DP.
In the central DP, the domain of secrets  consists of datasets, and the adjacency relation $x\sim x'$ holds if and only if $x$ and $x'$ differ for one record.
In the local DP, the domain of secrets  consists of all possible values  of a record,  and the adjacency relation $x\sim x'$ holds for all    
values $x$ and $x'$ different from each other.

\begin{Definition}[Differential-privacy level of a channel]\rm\label{def:Vdp}
Let $C : \calx \times \caly \rightarrow \reals$ be a channel and let  $\sim$ be a binary symmetric relation on $\calx$.
We say that  $C $ is \emph{differentially private} if it is  \emph{conforming} to  $\sim$,  and, in this case,  
its \emph{differential-privacy level} $\vdp[C]$  is defined as
\begin{align*}
\vdp[C] = \max_{\substack{y, x, x': \\ x \sim x', \\ C(x,y)>0}}
 \ln \frac{C(x, y)}{C(x', y)}.
\end{align*}
\end{Definition}

As shown in~\cite{Dwork:14:Algorithmic}, the relation with the standard notion of differential privacy (cf. Definition~\ref{def:dp}) is that a channel $C$ provides $\varepsilon$-differential privacy iff  $\vdp[C] \le \varepsilon$. In other words, $\vdp[C]$ is the least value $\varepsilon$ for which $C$ is $\varepsilon$-differentially private. 
This presentation eases mathematical treatment, and
	effectively means that:
	(i) $0 \le \vdp[C] < \infty$; and 
	(ii) $\vdp[C] = \infty$ if and only if $C$ is not conforming to $\sim$. 
	Moreover, as usual, it is the case that the higher $\vdp[C]$, the less private channel $C$ is.

We will use  the level of privacy (or more appropriately, non-privacy, or leakage) $\vdp$ as the  definition of  utility in \edp-games. In the next sections, we will see that the properties of $\vdp$ are rather different from those of vulnerability in \qif-games.

\subsection{\edp-games vs. \qif-games}
\label{sub:compare:QIF:DP}

Intuitively, DP and QIF have a similar ``philosophy'': in both cases the attacker is trying to infer information about the secrets from the observables.\footnote{The interpretation of DP in terms of inference can be found, for instance, in ~\cite{Alvim:15:JCS,Barthe:11:CSF,Chatzikokolakis:13:PETS,Dwork:06:TCC,Kifer:14:TDS}.}
The notion of attacker in  \qif-games and \edp-games differ significantly. In both cases, the attacker is trying to gain some knowledge about the secrets from the observables, but the modalities are very different. Specifically:
\begin{itemize}
\item[(A)] In \edp-games, the prior does not appear in the definition of the payoff like it does in \qif-games. 
In \edp-games, the payoff is simply the minimal $\varepsilon$ such that $\varepsilon$-DP holds. 
\item[(B)] In \edp-games, the attacker is not trying to maximize her expected posterior probability of success on all observables like in \qif-games, but rather the maximum distinguishability (i.e., the ratio of the likelihoods) of two secrets for an observable whatsoever. 
This is what we call ``worst-case principle.''
\item[(C)] From the game-theoretic point of view, this
worst-case principle makes a big difference. In fact, we show  that it implies that the leakage is
not a convex function as in \qif-games, but only quasi-convex. Furthermore, even in the case of
visible choice, the leakage is not linear as in \qif-games, but only quasi-max, which is a particular
case of quasi-convexity.
\end{itemize}

\subsection{Formal definition of \edp-games }
Recall that for \qif-games
we defined (cf. \eqref{def:hidden-choice}) the hidden-choice composition 
$C_{\delta,a}$ resulting from a mixed strategy $\delta$ by the defender and a pure strategy $a$ by the attacker as:
\begin{align*}
C_{\delta a} \eqdef \expect_{d\leftarrow\delta} C_{da}~,
\end{align*}

In our \edp-games we will also consider the visible-choice composition  
modeling the case in which the attacker is able to identify the action taken by the defender.
This is because 
even in the case of visible choice, 
the level of privacy is not linear w.r.t. the channel composition. Hence 
visible-choice  \edp-games  are not captured by standard game theory.

In order to define the latter formally, let us introduce some notation.  Given a family $\{M_{i}\}_{i \in \cali}$ of
compatible matrices s.t. each $M_{i}$ has type 
$\calx \times \caly_{i} \rightarrow \reals$, their \emph{concatenation} $\bigconc_{i \in \cali}$ 
is the matrix having all columns of every matrix in the family,
in such a way that every column is tagged with the matrix it
came from.
Formally, 
$\left( \bigconc_{i \in \cali} M_{i} \right)(x,(y,j)) = \,M_{j}(x,y)$, 
if $y \in \caly_{j}$,
and the resulting matrix has type
$\calx \times \left( \bigsqcup_{i \in \cali} \caly_{i} \right) \rightarrow \reals$.\footnote{%
	For $\cali = \{ i_{1}, i_{2}, \ldots , i_{n} \}$,\,
	$\bigsqcup_{i \in \cali} \caly_{i} = \caly_{i_{1}} \sqcup \caly_{i_{2}} \sqcup \ldots \sqcup \caly_{i_{n}}$ denotes the \emph{disjoint union} 
	$\{ (y,i) \mid y \in \caly_{i}, i \in \cali \}$
	of the sets $\caly_{i_{1}}$, $\caly_{i_{2}}$, $\ldots$, $\caly_{i_{n}}$.
}
Fig.~\ref{fig:concatenation} provides an example of the 
concatenation of two matrices.

\begin{figure}[tb]
\centering
{\small
	$
	\begin{array}{|c|cc|}
		\hline
		M_{1} & y_{1} & y_{2} \\ \hline
		x_{1} & a & b \\
		x_{2} & c & d \\ \hline
	\end{array}
	$
	$\bigconc$
	$
	\begin{array}{|c|ccc|}
	\hline
	M_{2} & y_{1} & y_{2} & y_{3} \\ \hline
	x_{1} & e & f & g \\ 
	x_{2} & h & i & j \\ \hline
	\end{array} 
	~$~=~$~
	\begin{array}{|c|ccccc|}
	\hline
	M_{1} \conc M_{2} & (y_{1},1) & (y_{2},1) & (y_{1},2) & (y_{2},2) & (y_{3},2) \\ \hline
	x_{1} & a & b & e & f & g \\ 
	x_{2} & c & d & h & i & j \\ \hline
	\end{array}
	$
	}
\caption{Example of the concatenation of two matrices $M_{1}$ and $M_{2}$.}
\label{fig:concatenation}
\end{figure}

We now will define the \qm{visible choice} among
channel matrices, representing the situation
in which a channel is probabilistically picked from a
family of channels, and the choice of channel
is revealed to the attacker.
Formally, given  $\{C_{i}\}_{i \in \cali}$ of
compatible channels s.t. each $C_{i}$ has type 
$\calx \times \caly_{i} \rightarrow \reals$, 
and a distribution $\mu$ on $\cali$, their 
\emph{visible choice composition} $\VChoice{i}{\mu}$ 
is defined as 
\begin{align}
\label{def:visible-choice}
\VChoice{i}{\mu} C_{i} = \bigconc_{i \in \cali} \;\mu(i)\, C_{i}~.
\end{align}
It can be shown that the resulting matrix 
$\VChoice{i}{\mu} C_{i}$ has type 
$\calx \times \left( \bigsqcup_{i \in \cali} \caly_{i} \right) \rightarrow \reals$,
and it is also a channel~\cite{Alvim:18:POST}.
Fig.~\ref{fig:visible-choice} depicts the 
visible choice among two channel matrices $C_{1}$ and $C_{2}$, with probability
$\nicefrac{1}{3}$ and $\nicefrac{2}{3}$, respectively.

It is important to notice that both the result of a 
	hidden choice composition as in~\eqref{def:hidden-choice} and 
	of a visible choice composition as in~\eqref{def:visible-choice} 
	cannot increase the maximum level of differential privacy
	provided by the operand channels, as formalized in Theorem~\ref{thm:q-convex-Vdp} ahead.
	Intuitively, that is a consequence of the fact that
	both operators produce a new channel matrix only by concatenating, scaling and summing up columns of the original channel matrices,
	and these three transformations cannot increase the maximum ratio 
	between any two elements in a same column of the resulting matrix.

\begin{figure}
\centering
$
{\displaystyle\frac{1}{3}}\;\;
\begin{array}{|c|cc|}
\hline
C_{1} & y_{1} & y_{2} \\ \hline
x_{1} & \nicefrac{1}{4} & \nicefrac{3}{4} \\
x_{2} & \nicefrac{1}{2} & \nicefrac{1}{2} \\ \hline
\end{array}
$
	$\bigconc$
$
{\displaystyle\frac{2}{3}}\;\;
\begin{array}{|c|cc|}
\hline
C_{2} & y_{1} & y_{3} \\ \hline
x_{1} & \nicefrac{1}{2} & \nicefrac{1}{2} \\
x_{2} & \nicefrac{2}{3} & \nicefrac{1}{3} \\ \hline
\end{array}
~$~=~$~
\begin{array}{|c|cccc|}
\hline
\VChoice{i}{\{\nicefrac{1}{3},\nicefrac{2}{3}\}} C_{i}  & (y_{1},1) & (y_{2},1) & (y_{1},2) & (y_{3},2) \\ \hline
x_{1} & \nicefrac{1}{12} & \nicefrac{1}{4} & \nicefrac{1}{3} & \nicefrac{1}{3} \\ 
x_{2} & \nicefrac{1}{6} & \nicefrac{1}{6} & \nicefrac{4}{9} & \nicefrac{2}{9} \\ \hline
\end{array}
$
\caption{Example of visible choice among two channel matrices $C_{1}$ and $C_{2}$.}
\label{fig:visible-choice}
\end{figure}

We are now ready to formalize the concept of \edp-games.

\begin{Definition}[\edp-game]\rm
\label{dp-game}
A \emph{\edp-game} is a 
simultaneous game
$(\cald, \cala, C)$ where 
$\cald, \cala$ are the finite sets of actions of the attacker and the defender, respectively, and $C=\{C_{da}\}_{da}$ is a family of 
compatible channel matrices, all conforming to a symmetric binary relation 
$\sim$ on their inputs, indexed on pairs of actions  $d\in\cald,a\in\cala$.
The game is zero-sum, and the utility for 
the attacker of a  pure strategy profile $(d,a)$ is given by 
$\vdp[C_{d a}]$. 

When the defender's choice is hidden from the 
attacker, the attacker's utility of a mixed strategy profile $(\delta,\alpha)$ is defined as
\[
\vdp(\delta,\alpha)\eqdef \vdp\left[\VChoice{a}{\alpha}\,C_{\delta a}\right]~,
\]
where
$C_{\delta a} \eqdef \expect_{d\leftarrow\delta} \,C_{da}$
is the hidden choice as defined in~\eqref{def:hidden-choice}.

When the defender's choice is visible to the 
attacker, the attacker's utility of a mixed strategy profile $(\delta,\alpha)$ is defined as
\[
\vdp(\delta,\alpha)\eqdef \vdp\left[\VChoice{a}{\alpha}\,\VChoice{d}{\delta}\,C_{d a}\right]~.
\]
\end{Definition}

Note that in our definition of \edp-games, the defender
	is interested in minimizing maximum leakage, whereas
	the adversary is interested in maximizing this same quantity.
	Note that it is reasonable for the defender  
	always to try to minimize maximum leakage, as the
	guarantees of differential privacy are intended to hold
	independently of the adversary.
	However, it would be reasonable	to consider scenarios in 
	which the adversary would not be necessarily 
	interested in maximizing the 
	channel's maximum leakage, e.g., because 
	he adopts a different notion of relation $\sim$ than that 
	used by the defender.
	In such a case, the resulting game would not be, in general,
	zero-sum.
	A study of such games, however, is beyond the scope 
	of this paper, and we therefore assume here that defender 
	and attacker agree on the utility measure to be minimized/maximized.

Now we present an example of a \edp-game and show that $\vdp$ is \emph{not} convex w.r.t. hidden choice.

\begin{figure}[tb]
\begin{tabular}{cc}
\begin{minipage}{0.56\textwidth}
\begin{center}
{\small 
\begin{tabular}{c c c}
&
$a=0$
&
$a=1$
\\[-1ex]
\qquad
\\[-1ex]
\!$d = 0$
&
$\begin{array}{|c|c|c|}\hline
C_{00} & \!y=y_0\! & \!y=y_1\! \\ \hline
\!x=x_0\!    & 0.90 & 0.10 \\
\!x=x_1\!    & 0.10 & 0.90 \\ \hline
\end{array}
$
&
$\begin{array}{|c|c|c|}
\hline
C_{01} & \!y=y_0\! & \!y=y_1\! \\ \hline
\!x=x_0\!    & 0.01 & 0.99  \\
\!x=x_1\!    & 0.03 & 0.97   \\ \hline
\end{array}
$
\\
\qquad
\\
\!$d = 1$
&
$
\begin{array}{|c|c|c|}
\hline
C_{10} & \!y=y_0\! & \!y=y_1\! \\ \hline
\!x=x_0\!    & 0.01 & 0.99 \\
\!x=x_1\!    & 0.03 & 0.97 \\ \hline
\end{array}
$
\quad
&
$\begin{array}{|c|c|c|}
\hline
C_{11} & \!y=y_0\! & \!y=y_1\! \\ \hline
\!x=x_0\!    & 0.90 & 0.10 \\
\!x=x_1\!    & 0.10 & 0.90 \\ \hline
\end{array}
$
\end{tabular}
}
\end{center}
\caption{Four channels $C_{da}$ representing all possible behaviors of a \edp-game.}
\label{fig:DP-game-channels}
\end{minipage}
\hfill
\begin{minipage}{0.03\textwidth}
~
\end{minipage}
\hfill
\begin{minipage}{0.39\textwidth}
	\centering
	\includegraphics[width=0.9\columnwidth]{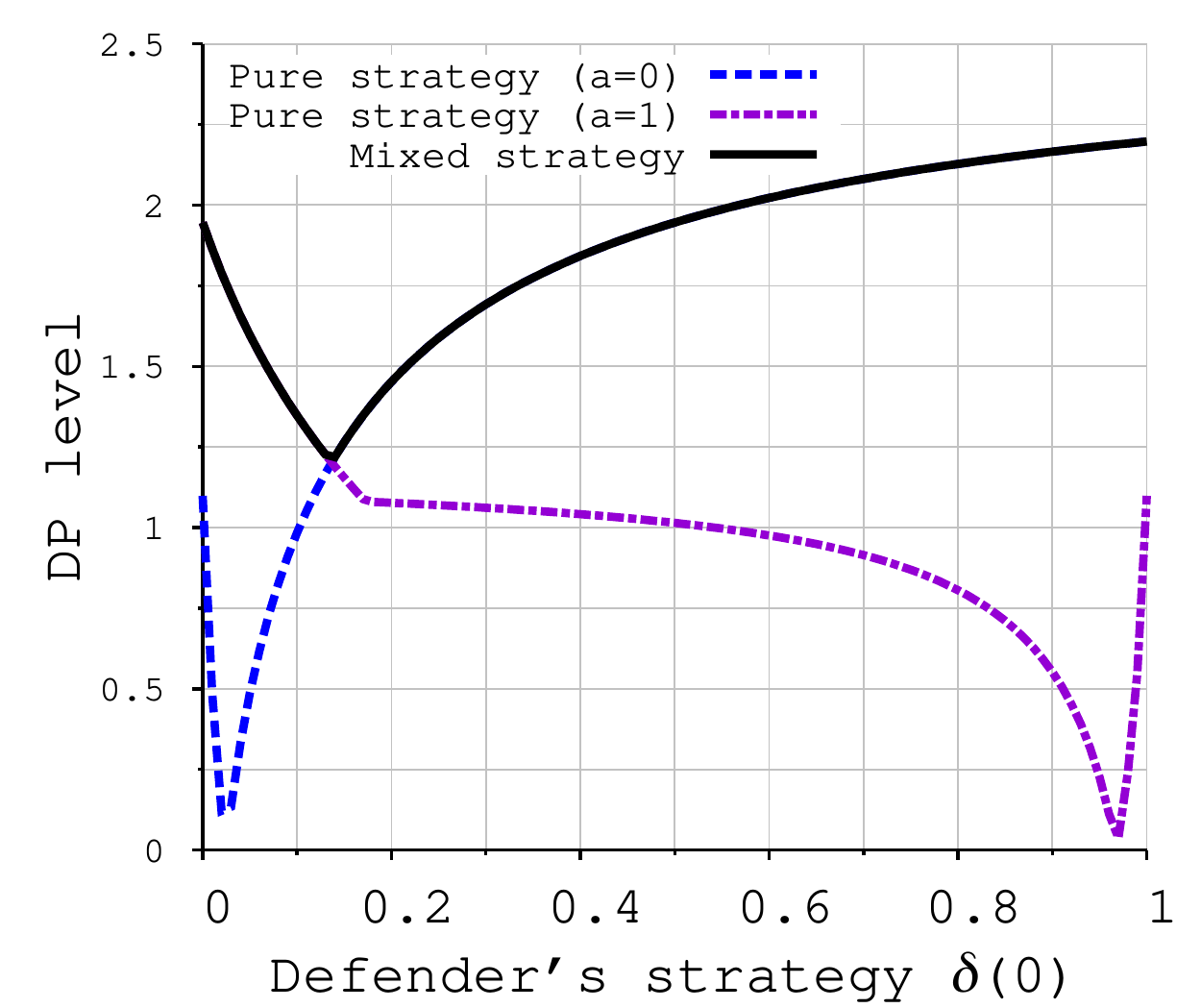}
	\caption{Relationship between $\delta$ and $\vdp$ under the attacker's optimal strategy $\alpha^*$.}
	\label{fig:dp-game}
\end{minipage}
\end{tabular}
\end{figure}

\begin{Example}[$\vdp$ is not convex w.r.t. hidden choice] \label{eg:dp-game}
Let us consider a \edp-game comprised of the action sets $\cald = \{0, 1\}$, $\cala = \{0, 1\}$, and four channels $C_{da}$ shown in \Fig{fig:DP-game-channels} over an input domain $\calx=\{x_0, x_1\}$ and an output domain $\caly=\{y_0, y_1\}$.
Assume also that $x_{0} \sim x_{1}$.
It is easy to see that the differential-privacy 
levels
of these
channels are $\vdp[C_{00}]=\vdp[C_{11}]= \ln{\max\{\nicefrac{0.90}{0.10},\nicefrac{0.90}{0.10}\}}=2.197$ and  $\vdp[C_{01}]=\vdp[C_{10}]= \ln{\max\{\nicefrac{0.03}{0.01},\nicefrac{0.99}{0.97}\}}=1.099$.
Now suppose that the attacker chooses $a=0$, and that the defender chooses either $d=0$ or $d=1$ with probability $\nicefrac{1}{2}$.
By Definition~\ref{def:Vdp},
\begin{align*}
\vdp[{\textstyle\nicefrac{1}{2}\,C_{0 0} + \nicefrac{1}{2}\,C_{0 1}}] &=
\textstyle
\max\bigl\{\, \ln\nicefrac{0.455}{0.065},\, \ln\nicefrac{0.935}{0.545} \,\bigr\} =
1.946 \\
\textstyle\nicefrac{1}{2}\,\vdp[C_{0 0}] + \nicefrac{1}{2}\,\vdp[C_{0 1}] &=
\textstyle\nicefrac{1}{2} \cdot 2.197 + \nicefrac{1}{2} \cdot 1.099 = 
1.648
{.}
\end{align*}
Hence we obtain 
$\vdp[{\textstyle\sum_{d} \nicefrac{1}{2}\,C_{0 d}}] = 1.946 >
1.648 = \sum_{d} {\textstyle\nicefrac{1}{2}}\,\vdp[C_{0 d}]$,
which implies that $\vdp$ is \emph{not} a convex function w.r.t. the defender's hidden choice.
\end{Example}

\subsection{Existence of Nash equilibrium for \edp-games with hidden choice}
\label{sec:dp-games-quasi-convexity}
We first show that the utility functions of \edp-games are quasi-convex w.r.t. hidden choice   composition, and quasi-max w.r.t. visible choice
(the term quasi-max is just our way of indicating the property expressed in (2) of the following theorem).

\begin{restatable}[$\vdp$ is quasi-convex/quasi-max on channel composition]{thm}{Convexmax}
\label{thm:q-convex-Vdp}
Let $\{C_{i}\}_{i \in \cali}$ be a family of compatible channels such that each $C_{i}$ is conforming to a symmetric binary relation $\sim$ over 
their input set, and
$\mu$ be a probability distribution on $\cali$.
Then:
\begin{enumerate}
\item\label{enumerate:vdp:quasi-convex}
$\vdp$ is quasi-convex w.r.t. hidden choice: 
$\displaystyle
\vdp\bigl[\expect_{i\leftarrow\mu} C_{i} \bigr] \leq \max_{i\in\supp{\mu}} \vdp[C_{i}]~.
$
\item\label{enumerate:vdp:quasi-max}
$\vdp$ is quasi-max w.r.t. visible choice:
$\displaystyle
\vdp\bigl[ \VChoice{i}{\mu}C_i \bigr] = \max_{i\in\supp{\mu}} \vdp[C_i]~.
$
\end{enumerate}
\end{restatable}

We are now ready to provide an upper bound on the utility under any strategy in a \edp-game with hidden choice.

\begin{restatable}[Upper bound on the utility for hidden choice]{propo}{bound}
\label{prop:upper:hidden}
A \edp-game provides $(\max_{d,a}\, \vdp(d,a))$-differential privacy:
for any mixed strategy profile $(\delta,\alpha)$,
\[
\vdp\Bigl[\VChoice{a}{\alpha}\, \displaystyle\expect_{d\leftarrow\delta} C_{da} \Bigr] \leq 
\max_{\substack{d\in\supp{\delta}\\ a\in\supp{\alpha}}}\, \vdp[C_{da}].
\]
\end{restatable}

Now we prove the existence of a Nash equilibrium in \edp-games as follows.
Interestingly, there is an optimal strategy for the attacker that is independent of the defender's strategy due to the fact that the attacker's strategy is visible to himself.
\begin{restatable}[Nash-equilibria for \edp-games with hidden choice]{thm}{NashEquilibrium}
\label{prop:Nash:DP}
For any \edp-game with hidden choice, there exists a Nash equilibrium.
Moreover, an arbitrary mixed strategy $\alpha^*$ such that $\supp{\alpha^*} = \cala$ is an optimal strategy for the attacker.
\end{restatable}

\subsection{Computing equilibria for \edp-games with hidden choice}
\label{sec:dp-games-computation}
In this section, we show how to compute optimal strategies in \edp-games with hidden choice of the defender's action.
Recall that by Theorem~\ref{prop:Nash:DP}, an optimal strategy for the attacker is an arbitrary mixed strategy $\alpha^*$ such that $\supp{\alpha^*} = \cala$.
To compute an optimal strategy for the defender, we use the following result.

\begin{restatable}[Optimal strategy for \edp-games with hidden choice]{propo}{StrategyHiddenChoice}
\label{prop:opt:DP:hidden}
For any \edp-game with hidden choice, 
an optimal strategy for the defender can be obtained by linear programming.
\end{restatable}

\begin{Example}[Optimal strategy for a $\edp$-game with hidden choice] \label{eg:dp-game-Nash}
We use the \edp-game with hidden choice in Example~\ref{eg:dp-game} to illustrate the differential-privacy level $\vdp$ when changing the defender's mixed strategy $\delta$ in Fig.~\ref{fig:dp-game}.
Let $\alpha^*$ be an arbitrary mixed strategy such that $\supp{\alpha^*} = \cala$.
Then the level of $\vdp$ for $\alpha^*$ is the maximum of those for the pure strategies $a = 0, 1$;
hence $\alpha^*$ is an optimal strategy for the attacker.
The defender's optimal strategy $\delta^*$ is given by $\delta^*(0) \approx 0.14$ and $\delta^*(1) \approx 0.86$.
\end{Example}

\subsection{Nash equilibria for \edp-games with visible choice}
\label{sec:dp-games-visible-choice}

Analogously, we obtain the following theorem under a visible choice of the defender's action.

\begin{restatable}[Nash equilibrium for \edp-games with visible choice]{thm}{StrategyVisibleChoice}
\label{prop:opt:DP:visible}
For any \edp-game with visible choice, there exists a Nash equilibrium.
Moreover, an arbitrary mixed strategy $\alpha^*$ such that $\supp{\alpha^*} = \cala$ is an optimal strategy for the attacker, and a pure strategy $d^* \in \argmin_d \max_a(\vdp[C_{da}])$ is an optimal strategy for the defender.
\end{restatable}

\begin{Example}[Optimal strategy for a $\edp$-game with visible choice] \label{eg:dp-game-Nash-visible}
To illustrate the Nash equilibria for \edp-games with visible choice, we again use the \edp-game in Example~\ref{eg:dp-game}.
By Definition~\ref{def:Vdp}, 
$\vdp[C_{00}] \approx 2.20$, $\vdp[C_{01}] \approx 1.10$, $\vdp[C_{10}] \approx 1.10$, and $\vdp[C_{11}] \approx 1.95$.
Hence the defender's optimal strategy is the pure strategy $d^* = \argmin_d \max_{a} \vdp[C_{da}] = 1$.
\end{Example}

Finally, we compare the \edp-games with visible choice with those with hidden choice in terms of $\vdp$ as follows.

\begin{restatable}[Visible choice $\geqslant$ hidden choice]{propo}{OrderDPVisibleHidden}
\label{prop:order:dp-games}
The \edp-games with hidden choice of the defender's action do not leak more than those with visible choice, i.e.,
\begin{align*}
\min_\delta\max_\alpha \vdp\Bigl[\VChoiceDouble{d}{\delta}{a}{\alpha} C_{da}\Bigr]
\;\; \geq\;\;
\min_\delta\max_\alpha\vdp\Bigl[\VChoice{a}{\alpha}\, \expect_{d\leftarrow\delta}\, {C_{d a}}\Bigr]
{.}
\end{align*}
\end{restatable}

\section{Information-leakage games vs. standard game theory models}
\label{sec:comparison-standart-gt}

In this section, we elaborate on the  differences between our information-leakage games and standard approaches to game theory.
In particular, we discuss:
(1) why the use of vulnerability as a utility function  makes 
\qif-games and \edp-games non-standard w.r.t. the von Neumann-Morgenstern's treatment of 
utility;
(2) why the use of concave utility functions to model risk-averse players
does not capture the behavior of the attacker in \qif-games; and
(3) how \qif-games differ from traditional convex-concave games.

\subsection{The von Neumann-Morgenstern's treatment of utility}

In their treatment of utility, 
von Neumann and Morgenstern~\cite{VonNeumann:47:Book}
demonstrated that the utility of a mixed strategy equals the expected utility 
of the corresponding pure strategies when a set of axioms is satisfied for 
player's preferences over probability distributions (a.k.a. \emph{lotteries}) on payoffs.
Since in our information-leakage games the utility of a mixed strategy is \emph{not} the
expected utility of the corresponding pure strategies, it is relevant to identify how
exactly our framework fails to meet von Neumann and Morgenstern (vNM) axioms.

Let us first introduce some notation.
Given two mixed strategies $\sigma$, $\sigma'$ for a player, we write
$\sigma \preceq \sigma'$ (or $\sigma' \succeq \sigma$) when the player 
prefers $\sigma'$ over $\sigma$, and $\sigma \sim \sigma'$
when the player is indifferent between $\sigma$ and $\sigma'$.
Then, the vNM axioms can be formulated as follows~\cite{Rubinstein:12:Book}.
For every mixed strategies $\sigma$, $\sigma'$ and $\sigma''$:
\begin{description}
\item[A1] \emph{Completeness}: it is either the case that
$\sigma \preceq \sigma'$, $\sigma \succeq \sigma'$, or $\sigma \sim \sigma'$.

\item[A2] \emph{Transitivity}: if $\sigma \preceq \sigma'$ and $\sigma' \preceq \sigma''$, 
then $\sigma \preceq \sigma''$.

\item[A3] \emph{Continuity}: if $\sigma \preceq\sigma'\preceq\sigma''$, then 
there exist $p \in [0,1]$ s.t. $p\,\sigma{+}(1-p)\,\sigma'' \sim\sigma'$.

\item[A4] \emph{Independence}: if $\sigma \preceq \sigma'$ then for any $\sigma''$
and   $p \in [0,1]$ we have $p\,\sigma + (1-p)\,\sigma'' \preceq p\,\sigma' + (1-p)\,\sigma''$.
\end{description}

The utility function $\vf[\pi,C]$ for \qif-games (for any fixed prior $\pi$ on secrets),
and the utility function $\vdp[C]$ for \edp-games are both total functions on $\calc$ 
ranging over the reals, and therefore satisfy axioms A1, A2 and A3 above.
However, neither satisfy A4, as the next example illustrates.
\begin{Example}
Consider the following three channel matrices from input set $\calx = \{0,1\}$
to output set $\caly=\{0,1\}$, where $\delta$ is a small positive constant.
\begin{align*}
{\small 
	\begin{array}{|c|c|c|}
	\hline
	C_{1} & y = 0 & y = 1 \\ \hline
	x = 0 & 1-2\delta & 2\delta \\ 
	x = 1 & 2\delta & 1-2\delta \\ \hline
	\end{array}
	\qquad 
	\begin{array}{|c|c|c|}
	\hline
	C_{2} & y = 0 & y = 1 \\ \hline
	x = 0 & 1-\delta & \delta \\ 
	x = 1 & \delta & 1-\delta \\ \hline
	\end{array}
	\qquad
	\begin{array}{|c|c|c|}
	\hline
	C_{3} & y = 0 & y = 1 \\ \hline
	x = 0 & \delta & 1-\delta \\ 
	x = 1 & 1-\delta & \delta \\ \hline
	\end{array}
	}
\end{align*}

In a \qif-game, if we focus on posterior Bayes vulnerability
on a uniform prior $\pi_{u}$, it is clear that an attacker would prefer $C_{2}$ over $C_{1}$, because
$$
\postvbayes{\pi_{u}}{C_{1}} = 1 - 2\delta < 1 - \delta = \postvbayes{\pi_{u}}{C_{2}}~.
$$
Similarly, in a \edp-game the attacker would also
prefer $C_{2}$ over $C_{1}$, since 
$$
\vdp[C_{1}] = \nicefrac{(1-2\delta)}{2\delta} <
\nicefrac{(1-\delta)}{\delta} = \vdp[C_{2}]~.
$$
	
However, for the probability $p = \nicefrac{1}{2}$ we would have the following
hidden composition:
\begin{align*}
{\small 
	\begin{array}{|c|c|c|}
	\hline
	p\,C_{1} + (1-p)\,C_{3} & y = 0 & y = 1 \\ \hline
	x = 0 & \nicefrac{(1-\delta)}{2} & \nicefrac{(1+\delta)}{2} \\
	x = 1 & \nicefrac{(1+\delta)}{2} & \nicefrac{(1-\delta)}{2} \\ \hline
	\end{array} 
	\quad \text{and} \quad
	\begin{array}{|c|c|c|}
	\hline
	p\,C_{2} + (1-p)\,C_{3} & y = 0 & y = 1 \\ \hline
	x = 0 & \nicefrac{1}{2} & \nicefrac{1}{2} \\
	x = 1 & \nicefrac{1}{2} & \nicefrac{1}{2} \\ \hline
	\end{array}
	}
\end{align*}

But notice that the channel $p\,C_{1} + (1-p)\,C_{3}$ clearly reveals no less information about the secret than channel $p\,C_{2} + (1-p)\,C_{3}$, and the utility 
of the corresponding \qif-game would be
$$
\postvbayes{\pi_{u}}{p\,C_{1} + (1-p)\,C_{3}} = \nicefrac{(1+\delta)}{2} > \nicefrac{1}{2} = \postvbayes{\pi_{u}}{p\,C_{2} + (1-p)\,C_{3}}~.
$$
Similarly, for a \edp-game we would have
$$
\vdp[p\,C_{1} + (1-p)\,C_{3}] = \nicefrac{(1+\delta)}{(1-\delta)} > 1 = \vdp[p\,C_{2} + (1-p)\,C_{3}]~.
$$

Hence, in both kinds of games we have
$C_{1} \preceq C_{2}$, but 
$p\,C_{1} + (1-p)\,C_{3} \succeq p\,C_{2} + (1-p)\,C_{3}$, and the axiom of
independence is not satisfied.
\end{Example}

It is actually quite natural that neither vulnerability nor differential privacy satisfies 
independence: a convex combination of two \qm{informative} channels 
(i.e., high-utility outcomes) can produce a \qm{non-informative} channel 
(i.e., a low-utility outcome)
whereas we can never obtain a \qm{more informative} channel from a convex combination of \qm{non-informative} channels.
As a consequence, the traditional 
game-theoretic approach to the utility of mixed strategies does not 
apply to our information-leakage games.

\subsection{Risk functions}

At first glance, it may seem that our \qif-games could be expressed with some clever use of the concept of a
\emph{risk-averse player} (which, in our case, would be the attacker), 
which is also based on convex utility functions (cf.~\cite{Osborne:94:BOOK}).
There is, however, a crucial difference: in the models of risk-averse
players, the utility function is convex 
\emph{on the payoff of an outcome of the game}, 
but the utility of a mixed strategy is still  
\emph{the expectation of the utilities of the pure strategies}, i.e., it is linear on the distributions.
On the other hand, the utility of mixed strategies in our \qif-games is \emph{convex on the distribution of the defender}.
This difference arises precisely because in \qif-games utility
is defined as the vulnerability of the channel perceived by the
attacker, and, as we discussed, this creates an extra layer of
uncertainty for the attacker.

\subsection{Convex-concave games}

Another well-known model from standard game theory is that of convex-concave
games, in which each of two players can choose among a continuous set of 
actions yielding convex utility for one player, and concave for the other.
In this kind of game, the Nash equilibria are given by \emph{pure strategies} for 
each player. 

A natural question would be why not represent our systems
as convex-concave in which the pure actions of players are
the mixed strategies of our \qif-games.
Namely, the real values $p$ and $q$ that uniquely determine the defender's 
and the attacker's mixed strategies, respectively, in the 
two-millionaires game of Section~\ref{sec:qif-games-definition}, could be 
taken to be the choices of pure strategies in a convex-concave  game
in which the set of actions for each player is the real interval $[0,1]$. 

This mapping from \qif-games to convex-concave, however, would not
be natural.
One reason is that utility is still defined as expectation in the standard 
convex-concave, in contrast to \qif-games. 
Consider two strategies $p_{1}$ and $p_{2}$ with utilities $u_1$ and $u_2$, respectively.
If we mix them using the coefficient $q\in [0,1]$, 
the resulting strategy 
 $q\, p_{1} + (1-q)\, p_{2}$
 will have utility  $u=q\, u_{1} + (1-q)\, u_{2}$ in the standard convex-concave game, 
 while in our case the utility would in general be strictly smaller than $u$.
The second reason is that a pure action corresponding to a mixed strategy may not always be realizable.
To illustrate this point, consider again the two-millionaires game, and the defender's mixed strategy consisting in choosing 
Program $0$ with probability $p$ and Program $1$ with probability $1-p$.
The requirement that the defender has a pure action corresponding to  $p$ implies the existence of a program (on Jeeves' side) that internally makes a probabilistic choice with bias $p$ and, depending on the outcome, executes Program $0$ or  Program $1$. 
However, it is not granted that Jeeves disposes of such a program. Furthermore, Don would not know what choice has actually been made, and thus the program would not achieve the same functionality, i.e., let Don know who is the richest. (Note that Jeeves should not communicate to Don the result of the choice, because of the risk that Alice intercepts it.) 
This latter consideration underlines a key practical aspect of \qif-games,
namely, the defender's advantage over the attacker due to his knowledge 
of the result of his own random choice (in a mixed strategy). This advantage would be lost in a convex-concave representation of the game since the random choice would be ``frozen'' in its representation as a pure action.

\section{Case studies}
\label{sec:case-study}

\subsection{The Crowds protocol as a \qif-game}
\label{sub:case-study-qif}
In this section, we apply our game-theoretic analysis to the case of anonymous
communication on a mobile ad-hoc network (MANET). 
In such a network, nodes can move in space and communicate with other nearby nodes. 
We assume that nodes can also access some global 
(wide area) network, but such
connections are neither anonymous nor trusted. Consider, for instance, smartphone
users who can access the cellular network, but do not trust the network
provider. The goal is to send a message on the global network without revealing
the sender's identity to the provider. For that, users can form
a MANET using some short-range communication method (e.g., Bluetooth), and take
advantage of the local network to achieve anonymity on the global one.

Crowds~\cite{Reiter:98:TISS} is a protocol for anonymous communication that can
be employed on a MANET for this purpose.
Note that, although more advanced systems for anonymous communication
exist (e.g., Onion Routing), the simplicity of Crowds makes it particularly
appealing for MANETs.
The protocol works as follows:
the \emph{initiator} (i.e., the node who wants to send the message) selects
some other node connected to him (with uniform probability) and forwards the
request to him. A \emph{forwarder}, 
upon receiving the message, performs a
probabilistic choice: with probability $p_f$ he keeps forwarding the message
(again, by selecting uniformly a user among the ones connected to him), while
with probability $1-p_f$ he delivers 
the message on 
the global network. 
Replies,
if any, can be routed back to the initiator following the same path in reverse
order.

Anonymity comes from the fact that the \emph{detected} node (the last in the
path) is most likely not the initiator. Even if the attacker knows the network
topology, he can infer that the initiator is most likely a node close to the
detected one, but if there are enough nodes we can achieve some reasonable
anonymity guarantees. However, the attacker can gain an important advantage by
deploying a node himself and participating to the MANET. When a node forwards a
message to this \emph{corrupted} node, this action is observed by the attacker
and increases the probability {of the forwarding node being the initiator. Nevertheless,
the node can still claim that he was only forwarding the request for someone
else; hence we still provide some level of anonymity. By modeling the system as
a channel, and computing its posterior Bayes vulnerability
\cite{Smith:09:FOSSACS}, we get the probability that the attacker correctly guesses
the identity of the initiator, after performing his observation.
\begin{figure}[tb]
	\begin{tabular}{cc}
	\begin{minipage}{0.25\textwidth}
	\centering
	\includegraphics[width=0.90\columnwidth]{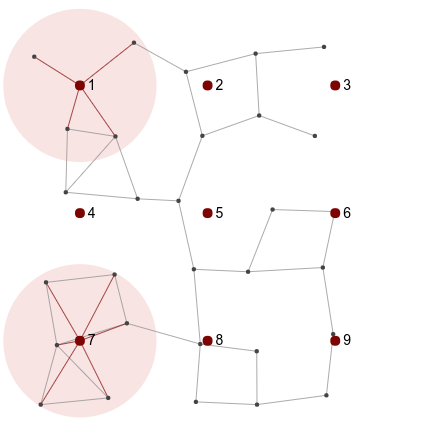}
	\caption{A MANET with 30 users in a $1$km$\times 1$km area.}
	\label{fig:manet}
	\end{minipage}
	\hfill
	\begin{minipage}{0.70\textwidth}
	\centering
	\tblcaption{Utility (Bayes vulnerability) for each pure strategy profile ($\%$).}
	\label{table:utilities}
	\begin{footnotesize}\vspace{-3ex}
		\[\def\arraystretch{0.95}
		\begin{array}{c|c|ccccccccc|}
		\multicolumn{2}{c}{}  & \multicolumn{9}{c}{\text{Attacker's action}} \\ \cline{2-11}
		&   & \textbf{1} & \textbf{2} & \textbf{3} & \textbf{4} & \textbf{5} & \textbf{6} & \textbf{7} & \textbf{8} & \textbf{9}  \\ \cline{2-11}
		\multirow{9}{*}{\rotatebox[origin=c]{90}{\text{Defender's action}}}
		& \textbf{1} & 7.38 & 6.88 & 6.45 & 6.23 & 7.92 & 6.45 & 9.32 & 7.11 & 6.45 \\
		& \textbf{2} & 9.47 & 6.12 & 6.39 & 6.29 & 7.93 & 6.45 & 9.32 & 7.11 & 6.45 \\
		& \textbf{3} & 9.50 & 6.84 & 5.46 & 6.29 & 7.94 & 6.45 & 9.32 & 7.11 & 6.45 \\
		& \textbf{4} & 9.44 & 6.92 & 6.45 & 5.60 & 7.73 & 6.45 & 9.03 & 7.11 & 6.45 \\
		& \textbf{5} & 9.48 & 6.91 & 6.45 & 6.09 & 6.90 & 6.13 & 9.32 & 6.92 & 6.44 \\
		& \textbf{6} & 9.50 & 6.92 & 6.45 & 6.29 & 7.61 & 5.67 & 9.32 & 7.11 & 6.24 \\
		& \textbf{7} & 9.50 & 6.92 & 6.45 & 5.97 & 7.94 & 6.45 & 7.84 & 7.10 & 6.45 \\
		& \textbf{8} & 9.50 & 6.92 & 6.45 & 6.29 & 7.75 & 6.45 & 9.32 & 6.24 & 6.45 \\
		& \textbf{9} & 9.50 & 6.92 & 6.45 & 6.29 & 7.92 & 6.24 & 9.32 & 7.11 & 5.68 \\\cline{2-11}
		\end{array}
		\]
	\end{footnotesize}
	\end{minipage}
	\end{tabular}
\end{figure}

In this section, we study a scenario of 30 nodes deployed in an area of
$1$km$\times 1$km, in the locations illustrated in Fig.~\ref{fig:manet}. Each
node can communicate with others up to a distance of 250 meters, forming the
network topology shown in the graph. To compromise the anonymity of the system,
the attacker plans to deploy a corrupted node in the network; the question is
which is the \emph{optimal location} for such a node. The answer is far from trivial:
on the one side being connected to many nodes is beneficial,  but at the same
time these nodes need to be ``vulnerable'', being close to a highly connected
clique might not be optimal.
At the same time, the administrator of the network is suspecting that the
attacker is about to deploy a corrupted node. Since this action cannot be
avoided (the network is ad-hoc), a countermeasure is to deploy a
\emph{deliverer} node at a location that is most vulnerable.
Such a node directly delivers all messages forwarded to it on the global
network; since it
never generates messages, its own anonymity is not an issue; it only
improves the anonymity of the other nodes. Moreover, since it never communicates
in the local network, its operation is invisible to the attacker.
But again, 
the optimal location for the new deliverer node is not obvious, and most importantly,
the choice depends on the choice of the attacker.

To answer these questions, we model the system as a \qif-game where the actions of 
attacker and defender are the locations of newly deployed corrupted
and honest nodes, respectively. We assume that the 
possible locations for new nodes are the nine ones shown in
Fig.~\ref{fig:manet}. For each pure strategy profile $(d,a)$, we construct the
corresponding network and use the PRISM model checker to construct the
corresponding channel $C_{da}$, using a model similar to the one
of~\cite{Shmatikov:02:CSFW}.
Note that the computation considers the specific network topology
of Fig.~\ref{fig:manet}, which reflects the positions of each node at the time
when the attack takes place; the corresponding channels need to be recomputed if
the network changes in the future.
As a leakage measure, we use the
posterior Bayes vulnerability (with uniform prior $\pi$), which is the
attacker's probability of correctly guessing the initiator given his
observation in the protocol. According to Definition~\ref{eq:v-mixed}, for a mixed
strategy profile $(\delta,\alpha)$ the utility is 
$\vf(\delta,\alpha)= \expectDouble{a\leftarrow\alpha}{}\hspace{-0.5ex}\postvf{\pi}{C_{\delta a}}$.

The utilities  (posterior Bayes vulnerability \%) for each pure profile are shown in Table~\ref{table:utilities}.
Note that the attacker and defender actions substantially affect the
effectiveness of the attack, with the probability of a correct guess ranging
between $5.46\%$ and $9.5\%$.
Based on 
Section~\ref{sec:qif-games-computation},
we can compute the best strategy for the defender, which turns out to be (probabilities expressed as \%):
\[
	\delta^* = (34.59,  3.48,  3.00,  10.52,  3.32,  2.99,  35.93,  3.19,  2.99)
\]
This strategy is part of an equilibrium 
and guarantees
that for any choice of the attacker the vulnerability is at most $8.76\%$, and
is substantially better than the best pure strategy (location~1) which leads to
the worst vulnerability of $9.32\%$. As expected, $\delta^*$ selects the most
vulnerable locations (1 and 7) with the highest probability. 
Still, the other locations are  selected with non-negligible probability, which is
important for maximizing the attacker's uncertainty about the defense.
Finally, note that for this specific $\delta^*$, the adversary's
best response is the pure strategy 1.

\subsection{Design of an LDP mechanism using a \edp-game}
\label{sub:case-study-dp}
In this section, we illustrate how to use our
	\edp-game framework in the design of a privacy mechanism. 
	We use a real dataset related to the COMPAS (Correctional Offender Management Profiling for Alternative Sanctions) risk assessment instrument developed by Equivant (former Northpointe).
	The tool provides several scores for individuals, including
	recidivism risk and violent recidivism risk,
	calculated using criminal history, jail and prison time, and demographics.
	The data we use in our case study is taken from the Propublica's study evaluating COMPAS' accuracy and bias
	on data from individuals from Broward County, in the USA, between the
	years of 2013 and 2014~\cite{compass:16:report}.

	In our case study, we consider that a defender is an individual
	who may participate in an independent study about criminal recidivism.
	The defender has access to his own COMPAS data, and is willing to disclose to a data curator the following non-sensitive attributes for the purpose of research: his \emph{ethnicity} ($z_1$), \emph{gender} ($z_2$), \emph{language} ($z_3$), and \emph{marital status} ($z_4$).
	On the other hand, he wishes to hide the status of his own \emph{agency\_text} attribute ($x$), taken from the COMPAS dataset, which is considered sensitive information.
	This attribute can take one of four values:
	\textit{BC} (Broward County),
	\textit{DRRD} (indicating that the individual is under the 
	Day Report Rentry Division, which helps
	reintegrate offenders back into the community following release from jail),
	\textit{pretrial},
	or
	\textit{probation}.
	However, the defender is aware that there are correlations
	between the non-sensitive values $z_1$, $z_2$, $z_3$, $z_4$ 
	and the sensitive value $x$, as shown in \Fig{fig:dp:correlation}.
	Therefore, the defender is interested in obfuscating his
	non-sensitive values before revealing them, in order
	to mitigate any possible leakage of information about $x$.

	To do that, the defender may employ a local DP mechanism 
	and add noise to the values of $z_1$, $z_2$, $z_3$, and $z_4$
	before reporting them to a data curator.
	On the other hand, we assume that a data analyst --here taking the role
	of the attacker-- is allowed to query the data curator
	about only one of the four attributes $z_1$, $z_2$, $z_3$, 
	or $z_4$.
	This assumption is justified by the fact that the combination of all
	four attributes could reveal too much information about $x$.
	We can capture this scenario with the following protocol:
\begin{enumerate}
\item The defender 
obfuscates his own record $(z_1, z_2, z_3, z_4)$ using 
one of four available privacy mechanisms 
$M_1, M_2, M_3, M_4$, and sends it to the data curator.
Each mechanism obfuscates the whole record $(z_1, z_2, z_3, z_4)$ 
at once, producing a new record $(z_1', z_2', z_3', z_4')$ of
randomized values to be reported to the data curator.
Hence, the defender's action $d$ is taken from 
the index set $\cald=\{1,2,3,4\}$ of the available mechanisms $M_1, M_2, M_3, M_4$.
We denote by $M_d$ the mechanism that the defender uses.
\item The data curator receives and stores the randomized record $(z_1', z_2', z_3', z_4')$ from the defender.
\item The attacker (i.e., the data analyst) can ask for the
data curator to reveal only one of the four sanitized values
$z_1'$, $z_2'$, $z_3'$ or $z_4'$,
and we denote the disclosed value by $y$.
The attacker's action $a$ is taken from the set $\cala=\{1,2,3,4\}$ of the attribute indices of $z_1'$, $z_2'$, $z_3'$, $z_4'$.
\end{enumerate}

\begin{figure}[tb]
	\begin{subfigure}{\linewidth}
		\centering
		\begin{footnotesize}
		\begin{tabular}{|c|cccccccc|}
			\hline
			& \textbf{African} & \textbf{Caucasian} & \textbf{Hispanic} & \textbf{Other} & \textbf{Arabic} & \textbf{Native} & \textbf{}Asian & \textbf{Oriental} 
			\\ \hline
			\textbf{BC} & 0.5366 & 0.3171 & 0.1463 & 0 & 0 & 0 & 0 & 0 \\
			\textbf{DRRD}  & 0.7234 & 0.1436 & 0.0957 & 0.0266 & 0.0053 & 0.0053 & 0 & 0 \\
			\textbf{Pretrial} & 0.4956 & 0.3509 & 0.0909 & 0.0545 & 0.0009 & 0.0034 & 0.0032 & 0.0004 \\
			\textbf{Probation} & 0.3267 & 0.3799 & 0.2590 & 0.0176 & 0.0017 & 0.0039 & 0.0101 & 0.0011 \\ \hline
		\end{tabular}
		\end{footnotesize}
		\caption{The conditional probabilities $p(z_1 \,|\, x)$ of ethnicity $z_1$ given sensitive information $x$.}
		\label{fig:dp:correlate:ethnic}
	\end{subfigure}
	\\
	\vspace{4mm}
	\begin{subfigure}{0.45\linewidth}
		\centering
		\begin{footnotesize}
		\begin{tabular}{|c|cc|}
			\hline
			& \textbf{Male} & \textbf{Female} \\ \hline
			\textbf{BC} & 0.731707 & 0.268293 \\
			\textbf{DRRD}  & 0.87234 & 0.12766 \\
			\textbf{Pretrial} & 0.802482 & 0.197518 \\
			\textbf{Probation} & 0.732053 & 0.267947 \\ \hline
		\end{tabular}
		\end{footnotesize}
		\caption{The conditional probabilities $p(z_2 \,|\, x)$ of gender $z_2$ given sensitive information $x$.}
		\label{fig:dp:correlate:gender}
	\end{subfigure}
	\hfill
	\begin{subfigure}{0.45\linewidth}
		\centering
		\begin{footnotesize}
		\begin{tabular}{|c|cc|}
			\hline
			& \textbf{English} & \textbf{Spanish} \\ \hline
			\textbf{BC} & 1.0000 & 0.0000 \\
			\textbf{DRRD}  & 1.0000 & 0.0000 \\
			\textbf{Pretrial} & 0.9969 & 0.0031\\
			\textbf{Probation} & 0.9935 & 0.0065 \\ \hline
		\end{tabular}
		\end{footnotesize}
		\caption{The conditional probabilities $p(z_3 \,|\, x)$ of language $z_3$ given sensitive information $x$.}
		\label{fig:dp:correlate:language}
	\end{subfigure}
	\\
	\vspace{4mm}
	\begin{subfigure}{\linewidth}
		\centering
		\begin{footnotesize}
		\begin{tabular}{|c|ccccccc|}
			\hline
			& \textbf{Single} & \textbf{Married} & \textbf{Divorced} & \textbf{Separated} & \textbf{Widowed} & \textbf{Other} & \textbf{Unknown} \\
			\hline
			\textbf{BC} & 0.9268 & 0.0000 & 0.0488 & 0.0244 & 0.0000 & 0.0000 & 	0.0000 \\
			\textbf{DRRD}  & 0.9149 & 0.0479 & 0.0106 & 0.0053 & 0.0053 & 0.0160 & 0.0000 \\
			\textbf{Pretrial} & 0.7664 & 0.1201 & 0.0482 & 0.0258 & 0.0050 & 0.0295 & 0.0050 \\
			\textbf{Probation} & 0.6820 & 0.1685 & 0.0990 & 0.0387 & 0.0094 & 0.0020 & 0.0003 \\ \hline
		\end{tabular}
		\end{footnotesize}
		\caption{The conditional probabilities $p(z_4 \,|\, x)$ of marital status $z_4$ given sensitive information $x$.}
		\label{fig:dp:correlate:marital}
	\end{subfigure}
	\caption{Correlation between the non-sensitive attributes $z_1$, $z_2$, $z_3$, $z_4$ and the sensitive attribute $x$. In each table, each row corresponds to a distribution on non-sensitive values $z_i$ given a particular sensitive value $x$.}
	\label{fig:dp:correlation}
\end{figure}

	We assume that when the defender selects a privacy mechanism, 
	he does not know which attribute $z_a$ the attacker will request. Symmetrically, the attacker does not know what privacy mechanism $M_d$ the defender uses.
	By using our framework of \edp-games, we design an optimal privacy mechanism by computing an optimal strategy for the defender.

Let $\calz_i$ be the range of the attribute $z_i$'s possible values, and 
$\calz = \prod_{i=1}^{4} \calz_i$.
To construct the defender's privacy mechanism, we define \emph{$(\epsilon, \calz_i)$-randomized response} $\RR{\epsilon}{\calz_i}: \calz_i\rightarrow\distr\calz_i$ by:
\begin{align*}
\RR{\epsilon}{\calz_i}(y \,|\, z) &=
\begin{cases}
\nicefrac{e^\varepsilon}{(|\calz_i| + e^\varepsilon - 1)}
& \text{if $y = z$} \\
\nicefrac{1}{(|\calz_i| + e^\varepsilon - 1)}
& \text{if $y \in \calz_i\setminus\{z\}$.}
\end{cases}
\end{align*}
Then, the defender's actions are realized by
mechanisms $M_d: \calz \rightarrow\distr\calz$  that apply $(0.1, \calz_d)$-randomized response $\RR{0.1}{\calz_d}$ to $z_d$ and $(2.0, \calz_j)$-randomized response $\RR{2.0}{\calz_j}$ to $z_j$ for all $j \neq d$.
Note that each mechanism $M_d$ adds significantly more noise
to attribute $z_d$ than to all other attributes.
For instance, $M_3(z_1, z_2, z_3, z_4) = (\RR{2.0}{\calz_1}(z_1), \RR{2.0}{\calz_2}(z_2), \RR{0.1}{\calz_3}(z_3), \RR{2.0}{\calz_4}(z_4))$.

To design an optimal privacy mechanism for the defender, we consider a $\edp$-game.
To set the game properly, we need to consider the \emph{full} mechanisms, namely those that operate on the secrets. Remember that the domain of secrets consists of the four possible values of the {\emph{agency\_text}} attribute (and 
the adjacency relation is the usual one for local differential privacy: $x\sim x'$ for all pairs of different values $x$ and $x'$). 
For each choice $a$ of one of the $z_i$'s attributes, and each choice $d$ of the mechanism $M_d$ that obfuscates the $z_i$'s, the corresponding full mechanism $C_{da}$ is obtained by 
composing the probabilistic relation from $x$ to $z_a$  with $M_d$. 
Note that this is similar to the 
addition of noise to the result of a query 
in standard differential privacy. 
The main difference is that a query is deterministic, while in our case the  relation between $x$ 
and $z_a$ is probabilistic. For this reason, the addition of the noise $M_d$ has to be done by multiplying the channel representing the probabilistic  relation and  $M_d$.
This kind of channel composition is called \emph{cascade}.

More formally: For a pure strategy profile $(d, a)\in\cald\times\cala$, the channel $C_{da}: \calx\times\calz_a\rightarrow\reals$ 
is the cascade composition of the channels representing the correlation $p(z_i \,|\, x)$ and the randomized response $p(y \,|\, z_i)$:
\begin{align*}
C_{da}(x, y) &= 
\begin{cases}
{\textstyle\sum_{z \in \calz_a}}\, \RR{0.1}{\calz_a}(y \,|\,  z)\, p(z \,|\, x)
& \text{if $d=a$} \\
{\textstyle\sum_{z \in \calz_a}}\, \RR{2.0}{\calz_a}(y \,|\,  z)\, p(z \,|\, x)
& \text{otherwise}.
\end{cases}
\end{align*}
Note that the output ranges over $\calz_a$, since the data curator provides the attacker with the data of only one of the four attributes.

For each channel $C_{da}$, the DP level $\vdp[ C_{da} ]$ is given by Definition~\ref{def:Vdp}.\!

When the defender's choice is hidden, we compute a solution for the \edp-game using the algorithm shown in Proposition~\ref{prop:opt:DP:hidden}.
We obtain the following optimal mixed strategy $\delta^* \in \distr\cald$ for the defender:
\[
\delta^* = (0.5714, 0.0183, 0.0000, 0.4103).
\]
Then the defender's optimal privacy mechanism $M^*: \calz\rightarrow\distr\calz$ is given by
$
M^* = \expect_{d\leftarrow\delta^*} [M_d],
$
and its differential privacy level is $\vdp[ M^* ] = 0.3892$.

By using $M^*$, each attribute is obfuscated by the following mechanism:
\begin{align*}
p(y \,|\, z_1) &= 0.5714 \,\RR{0.1}{\calz_1}(y \,|\, z_1) + (1 - 0.5714) \,\RR{2.0}{\calz_1}(y \,|\, z_1) \\
p(y \,|\, z_2) &= 0.0183 \,\RR{0.1}{\calz_2}(y \,|\, z_2) + (1 - 0.0183) \,\RR{2.0}{\calz_2}(y \,|\, z_2) \\
p(y \,|\, z_3) &= \RR{2.0}{\calz_3}(y \,|\, z_3) \\
p(y \,|\, z_4) &= 0.4103 \,\RR{0.1}{\calz_4}(y \,|\, z_4) + (1 - 0.4103) \,\RR{2.0}{\calz_4}(y \,|\, z_4).
\end{align*}
Then the optimal mechanism $M^*$ adds less perturbation to the attributes $z_2$ (gender) and $z_3$ (language).
This can be explained by observing $p(z_2 \,|\, x)$ and $p(z_3 \,|\, x)$ in \Figs{fig:dp:correlate:gender} and~\ref{fig:dp:correlate:language}.
Clearly, $z_2$ and $z_3$ leak little information on the secret $x$, so the defender need not add much noise to $z_2$ or $z_3$.

Note that to construct this optimal mechanism $M^*$, the defender needs to know the correlation $p(z_a \,|\, x)$ for each $a\in\cala$.
(This is reasonable in our scenario, since the defender 
has access to the original COMPAS dataset and can compute the correlations from it).
On the other hand, the attacker need not know the correlation $p(z_a \,|\, x)$ at all, since an arbitrary mixed strategy $\alpha^*$ with $\supp{\alpha^*}=\cala$ is optimal for the attacker.
(Similarly, note that this is reasonable in our scenario, and this
lack of knowledge is one of the reasons why the data analyst is
interested in collecting data from individuals in the COMPAS
dataset in the first place.)

When the defender's choice is visible, we compute a solution for the \edp-game using Theorem~\ref{prop:opt:DP:visible}.
The pure strategy $d = 4$ is optimal for the defender, while an arbitrary mixed strategy $\alpha^*$ with $\supp{\alpha^*}=\cala$ is optimal for the attacker.
The DP level of the equilibrium is given by $0.5994$.

\begin{table}[tb]
	\centering
	\begin{footnotesize}
		\[\def\arraystretch{1.15}
		\begin{array}{c|c|cccc|}
		\multicolumn{2}{c}{}  & \multicolumn{4}{c}{\text{Attacker's action}} \\ \cline{2-6}
		&   & \textbf{1} & \textbf{2} & \textbf{3} & \textbf{4}   \\ \cline{2-6}
		\multirow{3}{*}{\text{Defender's action}}
		& \textbf{1} & 0.0395 & 0.4020 & 0.0404 & 0.7306 \\
		& \textbf{2} & 0.5994 & 0.0145 & 0.0404 & 0.7306 \\
		& \textbf{3} & 0.5994 & 0.4020 & 0.0007 & 0.7306 \\
		& \textbf{4} & 0.5994 & 0.4020 & 0.0404 & 0.0237 \\\cline{2-6}
		\end{array}
		\]
	\end{footnotesize}
	\caption{Utility for each pure strategy profile, measured as $\vdp{}$.}
	\label{table:dp:channels}
\end{table}

\section{Conclusion and future work}
\label{sec:conclusion}

In this paper, we explored information-leakage  games,
in which a defender and an attacker have opposing goals in optimizing
the amount of information revealed by a system.
In particular, we discussed \qif-games, in which
utility is information leakage of a channel, and we introduced \edp-games, 
in which  utility is the level of differential privacy provided by the channel.
In contrast to standard game theory models, in our games the utility of 
a mixed strategy is either a convex, quasi-convex, or quasi-max function of the 
distribution of the defender's actions, rather than the expected value of  the  utilities of the pure strategies in the support.  
Nevertheless, the important properties of game theory, 
notably the existence of a Nash equilibrium, still hold for our zero-sum information-leakage games, 
and we provided algorithms to compute the corresponding optimal strategies 
for the attacker and the defender. 

As future research, we would like to extend information-leakage games to scenarios
with repeated observations, i.e., when the attacker can repeatedly observe 
the outcomes of the system in successive runs, under the assumption that 
both the attacker and the defender may change the channel at each run. 
This would incorporate the sequential and parallel compositions of quantitative information flow~\cite{Kawamoto:17:LMCS} in information-leakage games.
Furthermore, we would like to consider the possibility 
of adapting the defender's strategy to the secret value, as
we believe that in some cases this would provide a 
significant advantage to the defender. 
We would also like to deal with the cost of attack 
and of defense, which would lead to non-zero-sum games. 
Finally, we are also interested in incorporating our information leakage games in signaling games~\cite{Cho:87:QJE}, which model asymmetric information among players and are applied to economics and behavioral biology.

\begin{acks}
The authors are thankful to Arman Khouzani and Pedro O. S. Vaz de Melo 
for valuable discussions, 
and to Shintaro Miura for pointing out various related literature.
This work was supported by JSPS and Inria under the project LOGIS of the Japan-France AYAME Program,
by the PEPS 2018 project MAGIC, by the ANR project REPAS, 
and by the project Epistemic Interactive Concurrency (EPIC) from the STIC 
AmSud Program.
M\'{a}rio S. Alvim was supported by CNPq, CAPES, and FAPEMIG.
Yusuke Kawamoto was supported by ERATO HASUO Metamathematics for Systems Design Project (No. JPMJER1603), Japan,
and by JSPS KAKENHI Grant Numbers JP17K12667 and JP19H04113.
Catuscia Palamidessi was supported by the ERC grant HYPATIA under the European Union Horizon 2020 research and innovation programme.  
\end{acks}

\bibliographystyle{ACM-Reference-Format}
\bibliography{short,new,short-updated}

\appendix
\section{Proofs}
\label{sec:proofs}

\newenvironment{Reason}{\begin{tabbing}\hspace{2cm}\= \hspace{0.5cm} \=\hspace{3.5cm} \= \kill}
{\end{tabbing}\vspace{-1em}}
\newcommand\Step[2] {\>#1 \> $\begin{array}[t]{@{}llll}#2\end{array}$ \\[1ex]}
\newcommand\StepR[3] {\>#1 \> $\begin{array}[t]{@{}ll}#3\end{array}$ \> #2  \\[1ex]}
\newcommand\WideStepR[3] {#1 \>$\begin{array}[t]{@{}ll}~\\#3\end{array}$ \\[1ex]}
\newcommand\Space {~ \\}
\newcommand\RF {\small}

This appendix contains the proof of the formal results of this  paper.

\begin{proof}[\bf Proof for Corollary~\ref{cor:Nash}]
Given a mixed strategy profile $(\delta, \alpha)$, the utility $\vf(\delta, \alpha)$ given in Definition~\ref{eq:v-mixed}   is affine (hence concave) on $\alpha$.
Furthermore, by Theorem~\ref{theo:convex-V-q}, $\vf(\delta, \alpha)$ is convex on $\delta$. 
Hence we can apply the von Neumann's minimax
theorem (Section \ref{subsec:zero-sum-games}), which ensures the existence of a saddle point, i.e., a Nash equilibrium. 
\end{proof}

\begin{proof}[\bf Proof for Proposition~\ref{prop:minproblem}]
It is sufficient to show that for all mixed strategy $\delta$ for the defender: 
\[
\max_\alpha \vf(\delta,\alpha) = 
\max_a \,\postvf{\pi}{{\textstyle\expect_{d\leftarrow\delta}} C_{da}}.
\]

Let $\alpha$ be an arbitrary mixed strategy for the attacker.
Then we have that:
\begin{align*}
\vf(\delta,\alpha)
&=
\smallersum{a}\alpha(a) \,\postvf{\pi}{\expect_{d\leftarrow\delta}\, C_{da}}
\\ &\le
\smallersum{a}\alpha(a) (\max_a \,\postvf{\pi}{\expect_{d\leftarrow\delta}\,C_{da}})
& \text{(by $\alpha(a)\ge 0$)}
\\ &=
({\textstyle\max_a} \,\postvf{\pi}{\textstyle\expect_{d\leftarrow\delta}\,C_{da}})( \smallersum{a}\alpha(a)))
\\ &=
{\textstyle\max_a} \,\postvf{\pi}{\textstyle\expect_{d\leftarrow\delta}\,C_{da}}.
\end{align*}
This holds for any $\alpha$, hence $\max_\alpha \vf(\delta,\alpha) \le \max_a \,\postvf{\pi}{\expect_{d\leftarrow\delta}\,C_{da}}$;
the $(\ge)$ case is trivial since we can take $\alpha$ to be the point distribution on any $a$.
\end{proof}

\begin{proof}[\bf Proof for Theorem~\ref{prop:Lipschitz}]

	Our iterative method for computing
	the minimum value $f^* = \min_\delta f(\delta)$ is an application of the subgradient descent method, 
	that can be applied to any convex function. 
	
	We start by showing that $f(\delta)$ is convex. Let $\delta_1,\delta_2\in\distr\cald$, and $c\in[0,1]$ be a convex coefficient. 
	\begin{align*}
	f(c\delta_1 + (1-c) \delta_2)
	&=
	\max_a\postvf{\pi}{\smallersum{d}(c\delta_1(d) + (1-c) \delta_2(d))\,C_{da} }
	& \text{(by \eqref{eq:effe})}
	\\ &\le
	\max_a(c \postvf{\pi}{\smallersum{d} \delta_1(d)\,C_{da}} \, + \, (1-c)\postvf{\pi}{\smallersum{d}\delta_2(d)\,C_{da}})
	& \hspace{-3ex}\text{(by Theorem \ref{theo:convex-V-q})}
	\\ &\le
	c \max_a \postvf{\pi}{\smallersum{d} \delta_1(d)\,C_{da}} \, + \, (1-c)  \max_a \postvf{\pi}{\smallersum{d}\delta_2(d)\,C_{da}}
	\\ &=
	c f(\delta_1) + (1-c) f( \delta_2)\;.
	& \text{(by \eqref{eq:effe})}
	\end{align*}

	The proof of convergence of the subgradient decent method applied to $f$
	is provided by \cite[Sections~3 and 5]{Boyd:06:misc}.
	We only need to ensure that the two crucial conditions of this proof,
	stated in \cite[Section~3.1]{Boyd:06:misc}, are met.

	First, the proof requires that the norm of all subgradients $\|h^{(k)}\|_2$ is bounded
	by some constant $G$. For this, it is sufficient to show that $f(\delta)$ is
	Lipschitz; the latter follows from the fact that $\vf(\pi)$ itself is assumed to be Lipschitz.

	Second, the proof requires that there is a known number $R$ bounding the distance between
	the initial point $\delta^{(1)}$ and the optimal one, namely
	$\|\delta^{(1)} - \delta^*\|_2 \le R$. Since we start from the uniform distribution $u_\cald$, we take $R$
	to be the distance between $\delta^{(1)}$ and the point in the simplex maximally distant from it,
	which is a point distribution $\delta^{\text{point}} = (1,0,\ldots)$. That is, for $n = |\calx|$, we take:
	\[
		R = \|\delta^{(1)} - \delta^{\text{point}} \|_2
		= \sqrt{ \big(1 - \nicefrac{1}{n} \big)^2 + \big( 0 - \nicefrac{1}{n} \big)^2 (n-1) }
		= \sqrt{ \nicefrac{(n-1)}{n} } ~.
	\]
	
	Note also that the ``diminishing step size'' $s_k$ that we use is among those
	known to guarantee convergence \cite[Sections~3 and 5]{Boyd:06:misc}.

	The fact that $l^{(k)}$ is a lower bound on $f^*$ comes from \cite[Section~3.4]{Boyd:06:misc},
	using the $R$ above. As a consequence, the stopping criterion guarantees that
	\begin{equation}\label{eq44}
		f(\hat\delta) - f(\delta^*) \le \epsilon
		~.
	\end{equation}
	Let $\alpha$ be an arbitrary mixed strategy for the attacker.
	From \autoref{prop:minproblem} and the above:
	\begin{align*}
	\vf(\hat\delta, \alpha) - \epsilon & \le
	f(\hat\delta) - \epsilon
	& \text{(by def. of $f$ in \eqref{eq:effe})}
	\\ & \le
	f(\delta^*) & \text{(by \eqref{eq44})}
	\\ & =
	\vf(\delta^*,\alpha^*)
	& \text{(by \eqref{eq:effe})}
	\\[-0.5ex] & \le
	\vf(\hat\delta,\alpha^*). & \text{(since $(\delta^*,\alpha^*)$ is a saddle point)}
	\end{align*}
	Similarly for
		$\vf(\hat\delta,\alpha^*)
		\le
		\vf(\delta,\alpha^*) + \epsilon$,
	which concludes the proof.
\end{proof}

\begin{proof}[\bf Proof for Theorem~\ref{thm:q-convex-Vdp}]
(1) 
Since each $C_i$ is conforming to $\sim$, so is $\bigadd_{i} \mu(i)\,C_{i}$.
Then quasi-convexity of $\vdp$ w.r.t. hidden choice is derived as follows.
\begin{align*}
\vdp\bigl[\expect_{i\leftarrow\mu}\,C_{i} \bigr]
\,& = \max_{\substack{ y, x, x':~ x \sim x'\!, \\ \sum_{i}\mu(i)C_{i}(x,y) > 0}}\!  \ln 
{\textstyle\frac{ \sum_{i} \mu(i)\, C_{i}(x,y) }
{ \sum_{i} \mu(i)\, C_{i}(x',y) }} 
  & \text{(by def. of $\vdp$)}
\\[-1.0ex]
& \leq 
\max_{\substack{y, x, x':~ x \sim x'}}\,  \ln \max_{\substack{i\in\supp{\mu},\, \\ C_{i}(x,y) > 0}}
{\textstyle\frac{ \mu(i)C_{i}(x,y) }
{ \mu(i)C_{i}(x',y) } }
  & \text{($*$)}
\\[-0.6ex]
& = 
\max_{i\in\supp{\mu}}\, \max_{\substack{ y, x, x':~  x \sim x'\!, \\ C_{i}(x,y) > 0}}\,  \ln 
{\textstyle\frac{ C_{i}(x,y) }
{ C_{i}(x',y) } }
& \text{($\ln$ is monotone)}
\\
& = 
\max_{i\in\supp{\mu}} \vdp[C_i]
& \text{(by def. of $\vdp$)}
\end{align*}

To complete the proof, we derive the inequality ($*$)\footnote{In~\cite{Kawamoto:19:ESORICS,Kawamoto:19:Allerton}, a similar inequality is used to prove their main theorems.}.
Let $y\in\caly$ and $x, x'\in\calx$ such that $x\sim x'$ and $\sum_{i}\mu(i)C_{i}(x,y) > 0$.
Then there is an $i\in\supp{\mu}$ such that $C_{i}(x,y) > 0$.
Since $C$ is conforming to $\sim$ and $x\sim x'$, it holds for any $j$ that $C_{j}(x,y) > 0$ iff $C_{j}(x',y) > 0$.
Hence, for any $j\in\supp{\mu}$,
\begin{align*}
\mu(j)C_{j}(x,y) \le 
\mu(j)C_{j}(x',y)\cdot
\max_{\substack{i\in\supp{\mu},\\ C_{i}(x,y) > 0}}\, {\textstyle\frac{\mu(i)C_{i}(x,y)}{\mu(i)C_{i}(x',y)}}.
\end{align*}
Therefore, ($*$) follows from:
\begin{align*}
{\textstyle\sum_{j}}\, \mu(j)C_{j}(x,y) \le 
\max_{\substack{i\in\supp{\mu},\\ C_{i}(x,y) > 0}}\, {\textstyle\frac{\mu(i)C_{i}(x,y)}{\mu(i)C_{i}(x',y)} }\cdot
{\textstyle\sum_{j}}\, \mu(j)C_{j}(x',y).
\end{align*}
\\[0.5ex]
(2)
Let $C = \VChoice{i}{\mu}C_i$.
Since each $C_i$ is conforming to $\sim$, so is $C$.
Then the quasi-max property of $\vdp$ w.r.t. visible choice is derived as follows.
\begin{align*}
\vdp\bigl[ \VChoice{i}{\mu}C_i \bigr] 
\,&= \max_{\substack{i, y, x, x':~ x \sim x'\!,\\ C(x,(y,i))>0, \\ i \in \supp{\mu}}}
 \ln {\textstyle\frac{C(x, (y,i))}{C(x', (y,i))}}
&\hspace{-7ex} \text{(by def. of $\vdp$ and vis. choice)}
\\[-1.0ex]
&= \max_{\substack{i, y, x, x':~ x \sim x'\!,\\ C(x,(y,i))>0, \\ i \in \supp{\mu}}}
 \ln {\textstyle\frac{\mu(i)\,C_i(x, y)}{\mu(i)\,C_i(x', y)}}\hspace{-5ex} 
&\text{(by def. of visible choice)} 
\\[-0.5ex]
&= 
\max_{i\in\supp{\mu}} \vdp[C_i] & \text{(by def. of $\vdp$)}
\end{align*}
\end{proof}

\begin{proof}[\bf Proof for Proposition~\ref{prop:upper:hidden}]
The proof is based on the quasi-convexity and quasi-max of $\vdp$.
\begin{align*}
\vdp\bigl[\VChoice{a}{\alpha} C_{\delta a}\bigr] &= 
{\textstyle\max_{a\in\supp{\alpha}}} \vdp[C_{\delta a}]
& \left(\text{by Theorem~\ref{thm:q-convex-Vdp} (2)}\right)
\\[-1ex] &\leq
{\textstyle\max_{a\in\supp{\alpha}}} {\textstyle\max_{d\in\supp{\delta}}} \vdp[C_{da}]
{.}
& \left(\text{by Theorem~\ref{thm:q-convex-Vdp} (1)}\right)
\end{align*}
\end{proof}

\begin{proof}[\bf Proof for Theorem~\ref{prop:Nash:DP}]
By Theorem~\ref{thm:q-convex-Vdp} (2), it holds for any mixed strategy profile $(\delta,\alpha)$ that:
\[
	\vdp(\delta,\alpha)\eqdef \vdp\bigl[\VChoice{a}{\alpha}\,C_{\delta a}\bigr]
	= {\textstyle\max_{a\in\supp{\alpha}}}\, \vdp[C_{\delta a}]
	\leq {\textstyle\max_{a\in\cala}}\, \vdp[C_{\delta a}]~.
\]
Hence for any $\delta$, an arbitrary mixed strategy $\alpha^*$ s.t. $\supp{\alpha^*} = \cala$ maximizes $\vdp(\delta,\alpha)$.
Therefore:
\[
\textstyle
\min_{\delta} \max_{\alpha} \vdp(\delta, \alpha) = 
\min_{\delta} \vdp(\delta, \alpha^*) =
\max_{\alpha} \min_{\delta} \vdp(\delta, \alpha)
{,}
\]
i.e., there exists a Nash equilibrium.
\end{proof}

\begin{proof}[\bf Proof for Proposition~\ref{prop:opt:DP:hidden}]
Let $\alpha^*$ be an arbitrary mixed strategy such that $\supp{\alpha^*} = \cala$.
By Theorem~\ref{prop:Nash:DP}, 
\begin{align*}
\textstyle
\min_{\delta} \max_{\alpha} \vdp(\delta, \alpha) =&\, 
\textstyle
\max_{\alpha} \min_{\delta} \vdp(\delta, \alpha) \\
=&\, 
\textstyle
\min_{\delta} \vdp(\delta, \alpha^*) \\[-0.5ex]
=&\, 
\textstyle
\min_{\delta} \max_{a\in\cala} \vdp(\delta, a) \\[-1.0ex]
=&\, 
\textstyle
\min_{\delta} \max_{a\in\cala} \max_{x,x',y} 
	{\textstyle\frac{ \sum_{d} \delta(d) C_{da}(x,y) }{ \sum_{d} \delta(d) C_{da}(x,y) }}~.
\end{align*}
The above program is an instance of \emph{generalized fractional programming}\cite{barros2013discrete},
in which
we minimize the largest of a family of ratios of continuous functions:
\begin{align*}
\min_{\delta\in\dist\cald} \max_{j \in J}
	{\textstyle\frac{ f_j(\delta) }{ g_j(\delta) }}~,
	\quad\text{where }\quad~
	J &= \cala\times\calx\times\calx\times\caly ~,\\[-1.5ex]
	f_{j}(\delta) &= {\textstyle\smallsum_{d}}\, \delta(d) C_{da}(x,y)~, \\[-0.5ex]
	g_{j}(\delta) &= {\textstyle\smallsum_{d}}\, \delta(d) C_{da}(x',y)~, \qquad j=(a,x,x',y)\in J  ~.
\end{align*}
We can solve this program using the Dinkelbach-type algorithm \cite{barros2013discrete}
as follows. Define
\begin{align*}
	\lambda_k &= \max_{j \in J} {\textstyle\frac{ f_j(\delta_{k-1}) }{ g_j(\delta_{k-1}) }}~, \\[-0.5ex]
	F_k(\delta) &=   \max_{j \in J} \left[  f_j(\delta) - \lambda_k g_j(\delta) \right] ~, \\[-0.5ex]
	\delta_k &=   \argmin_{\delta\in\dist\cald } F_k(\delta) ~.
\end{align*}
We start from a uniform $\delta_0$, and iterate using the above formulas,
until $F_k(\delta_k) = 0$ for some $k \ge 1$, in which case $\delta_k$ is guaranteed to be the optimal
solution with value $\lambda_k$.
Note that for each iteration $k \ge 1$, $\lambda_k$ is constant (computed from $\delta_{k-1}$).
Moreover, the optimization problem for $\delta_k$, which we need to solve in every iteration,
requires to minimize the max of linear functions, and can be transformed into the following
\emph{linear} program:
\begin{align*}
	&\text{variables }~ \delta(d), z \\[-0,5ex]
	&\text{minimize }~ z \quad\quad\quad
	\text{s.t. }~
	1 = {\textstyle \smallsum_{d}}\, \delta(d)  ~\mbox{ and }~
	z  \ge f_j(\delta) - \lambda_k g_j(\delta) \quad \forall j\in J ~,
\end{align*}
since $f_j(\delta),g_j(\delta)$ are linear on $\delta$.
Therefore, an optimal strategy for the \edp-game is obtained by solving a sequence of linear programs.
\end{proof}

\begin{proof}[\bf Proof for Theorem~\ref{prop:opt:DP:visible}]
Similarly to Theorem~\ref{prop:Nash:DP},
an arbitrary mixed strategy $\alpha^*$ such that $\supp{\alpha^*} = \cala$ maximizes the attacker's payoff independently of $\delta$.
Hence, for any $\delta$ and $\alpha$,
\[
\vdp(\delta,\alpha) = {\textstyle\max_{d\in\supp{\delta}}\, \max_{a\in\cala}}\,\ \vdp[C_{da}].
\]
This is minimized for a point distribution $\delta^*$ s.t. $\delta^*(d^*) = 1$ for $d^* \in\argmin_d \max_{a} \vdp[C_{da}]$.
\end{proof}

\begin{proof}[\bf Proof for Proposition~\ref{prop:order:dp-games}]
For any $a\in \cala$ and any $\delta\in \distr\cald$ we have:
\begin{align}
\vdp[{\VChoice{d}{\delta}} {C_{d a}}]
 \,\, =&\,\,  {\textstyle\max_{d \in \supp{\delta}}} \,\vdp[C_{d a}]
& \text{(by Theorem~\ref{thm:q-convex-Vdp} (\ref{enumerate:vdp:quasi-max}))}
\nonumber
\\[-1ex] \geq&\,\, \,
\vdp[\expect_{d\leftarrow\delta}\, C_{d a}]
{.}
& \text{(by Theorem~\ref{thm:q-convex-Vdp} (\ref{enumerate:vdp:quasi-convex}))}
\label{eq:OrderDPVisibleHidden:1}
\end{align}
Then for any $\alpha\in \distr\cala$, any $\delta\in \distr\cald$, and any 
$a^\dag \in \argmax_{a\in\supp{\alpha}} \vdp[\expect_{d\leftarrow\delta}\, C_{d a}]$,
we obtain:
\begin{align*}
\vdp\biggl[\VChoiceDouble{d}{\delta}{a}{\alpha} C_{da}\biggr]
\; =&\;
{\textstyle\max_{a\in\supp{\alpha}}}
\vdp[{\VChoice{d}{\delta}} {C_{d a}}]
& \text{(by Theorem~\ref{thm:q-convex-Vdp} (\ref{enumerate:vdp:quasi-max}))}
\\[-2ex] \;\geq&\;
\vdp[{\VChoice{d}{\delta}} {C_{d a^\dag}}]
\\[-0.6ex] \;\geq&\;
\vdp[\expect_{d\leftarrow\delta}\, C_{d a^\dag}]
&\text{(by \eqref{eq:OrderDPVisibleHidden:1})}
\\[-1ex] \;=&\;
{\textstyle\max_{a\in\supp{\alpha}}}
\vdp[\expect_{d\leftarrow\delta}\, C_{d a}]
&\text{(by def. of $a^\dag$)}
\\[-0.5ex] \;=&\;
\vdp\bigl[\VChoice{a}{\alpha}\, \expect_{d\leftarrow\delta} {C_{d a}}\bigr].
& \text{(by Theorem~\ref{thm:q-convex-Vdp} (\ref{enumerate:vdp:quasi-max}))}
\end{align*}
Therefore the theorem follows immediately.
\end{proof}

\section{A Bayesian interpretation of differential privacy}
\label{sec:dp-bayes}
Differential privacy has been interpreted in terms of Bayesian inference~\cite{Alvim:15:JCS,Barthe:11:CSF,Chatzikokolakis:13:PETS,Dwork:06:TCC,Kifer:14:TDS}.
Here we present two alternative versions of this view.

In the following, we assume that a mechanism (or channel)
$C$ satisfies $\varepsilon$-differential privacy as
in Definition~\ref{def:dp}. 
The mechanism has an input set $\calx$, an output set $\caly$,
and is defined w.r.t. an adjacency relation $\sim$ on $\calx$.
Recall from Section~\ref{subsec:qif} that
given a prior distribution $\pi$ on inputs to the mechanism,
we can derive a joint distribution $p_{X,Y}$ on $\mathcal{X} \times \mathcal{Y}$ in the usual way, and from that marginal distributions
$p_{X}$ and $p_{Y}$, as well as conditional distributions
$p_{X \mid y}$ and $p_{Y \mid x}$ for particular non-zero 
values of $y \in \caly$ and $x \in \calx$.
When clear from the context, we may omit subscripts on probability distributions, writing, e.g., $p(y)$, $p(x,y)$, and
$p(x \mid y)$ for $p_{Y}(y)$, $p_{X,Y}(x,y)$, and
$p_{X \mid y}(x \mid y)$, respectively.

Using this notation, the $\varepsilon$-differential privacy property can be rewritten as:
\begin{equation}\label{eq:DP}
e^{-\epsilon}\leq \frac{p(y\mid x)}{p(y\mid x')}\leq e^\epsilon~, \qquad \qquad   \text{for all $y{\in}\caly, x,x'{\in}\calx$ s.t. $x \sim x'$.}
\end{equation}

\subsection{Interpretation of differential privacy in terms of hypothesis testing}
A first interpretation presents differential privacy in terms of a bound on the adversary's success in performing hypothesis testing on the secret value~\cite{Chatzikokolakis:13:PETS,Kifer:14:TDS}.
More precisely, it notes that differential privacy ensures that the adversary's a priori capability of 
distinguishing between two adjacent secrets (e.g., two neighboring datasets differing only by the presence/absence
of a single individual) 
does not change significantly after an 
observation of the mechanism's output (e.g., the answer to a query in that dataset).
This guarantee (which, of course, depends on the value of the parameter $\epsilon$) holds for every prior --and that is the reason DP is said to be \qm{independent from the adversary's knowledge.}

Formally, the guarantee that a mechanism satisfies
\begin{align}
	\label{eq:dp1}
	\frac{p(y \mid x)}{p(y \mid x')} &\leq e^{\epsilon}, & \text{for all $y{\in}\caly, x,x'{\in}\calx$ s.t. $x \sim x'$}~,
\end{align}
where $\sim$ is a symmetric adjacency relation, is equivalent to 
\begin{align}
	\label{eq:dp2}
	\frac{p(x \mid y)}{p(x' \mid y)} &\leq e^{\epsilon} \frac{p(x)}{p(x')}, & \text{for all $y{\in}\caly, x,x'{\in}\calx$ s.t. $x \sim x'$, and all priors $p(\cdot)$ on $\calx$}~.
\end{align}
	Note that in Equation~\eqref{eq:dp2}, the term $\nicefrac{p (x)}{p (x')}$ represents the adversary's
	capability of distinguishing between adjacent secrets $x$ and $x'$ a priori (i.e., before the output of the mechanism is observed), whereas the term $\nicefrac{p(x \mid y)}{p (x' \mid y)}$ represents his capability
	of distinguishing between the same adjacent secrets after 
	an output $y$ of the mechanism is observed.	
	Because of the universal quantification on all adjacent secret inputs $x,x'$, outputs $y$, and prior distribution $p(\cdot)$ on inputs, Equation~\eqref{eq:dp2} exactly states
	differential privacy's guarantees against maximum distinguishability of two secrets.

The proof of the equivalence of Equations \eqref{eq:dp1} and \eqref{eq:dp2} follows immediately by the Bayes Theorem, which we recall here:
\begin{align*}
	 &\text{Bayes Theorem} \,\,\,\,\,\,\,\, p(x \mid y) \,\,=\,\,\frac{p(y \mid x)~p(x)}{p(y)} \,.
\end{align*}
 Hence, we have: 
 \begin{align*}
	  \frac{p(x \mid y)}{p(x' \mid y)}  \,\, =  \,\, \frac{p(y \mid x)~p(x)}{p(y)}~\frac{p(y)}{(y \mid x')~p(x')} \,\, =  \,\, \frac{p(y\mid x)}{p(y \mid x')} \,\, \frac{p(x)}{p(x')}
	  \le e^{\epsilon} \frac{p(x)}{p(x')}.
\end{align*}

\subsection{Interpretation of differential privacy in terms of increase in information}
A second interpretation presents differential privacy in terms of a bound on the adversary's increase in information about
the secret value~\cite{Alvim:15:JCS,Barthe:11:CSF,Dwork:06:TCC}.
In this case, however, the exact formulation depends on the nature of the adjacency relation $\sim$. 
We present here the examples of local differential privacy~\cite{Kasiviswanathan:08:FOCS}, and of the standard differential privacy on datasets~\cite{Dwork:06:ICALP,Dwork:06:TCC}.

\subsubsection{Local differential privacy}
In local differential privacy, the relation $\sim$ holds between every two secrets values $x$ and $x'$. 
In this case,  the Bayesian formulation is expressed by the following bound, for some real number $\alpha\geq 0$:
\begin{equation}\label{eq:BI1}
e^{-\alpha}\leq \frac{p(x\mid y)}{p(x)}\leq e^\alpha~ \qquad \text{for all $y{\in}\caly, x {\in}\calx$, and all priors $p(\cdot)$ on $\calx$}~.
\end{equation}
Essentially, inequality \eqref{eq:BI1} 
means that the prior and posterior probabilities of the secret
taking a particular value 
do not differ too much. 
We now state precisely and prove the correspondence between DP and this Bayesian formulation. 

\begin{Theorem}\label{theo:correspondance}
	Given a certain mechanism $C$:	
	\begin{enumerate}
		\item If $C$  satisfies inequality \eqref{eq:DP}, then it satisfies inequality 
		\eqref{eq:BI1} 
		with $\alpha=\epsilon$.
		\item If $C$ satisfies inequality \eqref{eq:BI1},
		then it satisfies inequality \eqref{eq:DP} with $\epsilon=2\,\alpha$.
	\end{enumerate}
\end{Theorem}
\begin{proof}
	\begin{enumerate}
		\item Assume that  \eqref{eq:DP} is satisfied. Then:
		
		\begin{align*}
		\frac{p(x \mid y)}{p(x)} =&\,\,
		\frac{p(y \mid x)}{p(y)} & \text{(by the Bayes theorem)}\\[2ex]
		=&\,\, \frac{p(y  \mid x)}{\sum_{z\in\calx}p(y,z) } &\text{(marginals)}\\[2ex]
		=&\,\, \frac{p(y\mid x)}{\sum_{z\in\calx}p(y\mid z)p(z)} & \text{(chain rule)}\\[2ex]
		\leq & \,\, \frac{p(y\mid x)}{\sum_{z\in\calx}e^{-\epsilon} p(y\mid  x)p(z) } & \text{(by \eqref{eq:DP})}\\[2ex]
		=& \,\, \frac{p(y \mid x)}{e^{-\epsilon} p(y\mid x) } \\[2ex]
		=&\,\, e^{\epsilon}
		\end{align*}
		
		\item Assume that \eqref{eq:BI1} is satisfied. Then:

		\begin{align*}
		\frac{p(y\mid x)}{p(y \mid x')} =&\,\,
		\frac{p(x \mid y) \, p(y)}{p(x)} \,\frac{p(x' )}{p(x' \mid y)p(y)} & \mbox{(by the Bayes theorem)}\\[2ex]
		=& \,\, \frac{p(x \mid y)}{p(x)} \, \frac{p(x')}{p(x' \mid y)}\\[2ex]
		\leq & \,\, { e^{\alpha} \, e^{\alpha}}  &\mbox{  (by  \eqref{eq:BI1}}\\[2ex]
		=& \,\, { e^{2\, \alpha}}
		\end{align*}
	\end{enumerate}
\end{proof}

\subsubsection{Standard differential privacy on datasets }
Standard differential privacy is also called, nowadays, differential privacy in the central model. 
In this case, the relation $\sim$ holds between every two datasets  that differ for the presence or absence of a single record.
We will indicate by $x$ the presence of the target record and by $\neg x$ its absence. We will indicate by $s$ the rest of the dataset.  
By using this notation, the differential privacy  property can be rewritten as:
\begin{equation}\label{eq:DPcentral}
e^{-\epsilon}\leq \frac{P(y\mid x, s)}{P(y\mid \neg x,s)}\leq e^\epsilon~ \qquad \text{for all $y{\in}\caly, x {\in}\calx$, and datasets $s$} ~.
\end{equation}

The Bayesian formulation is expressed by the following bounds, for some real number $\alpha\geq 0$:
\begin{equation}\label{eq:BI3}
e^{-\alpha}\leq \frac{p(x\mid y, s)}{p(x\mid s )}\leq e^\alpha~ \qquad \text{for all $y{\in}\caly, x {\in}\calx$, datasets $s$, and all priors $p(\cdot)$ on $\calx$} ~.
\end{equation}
and
\begin{equation}\label{eq:BI4}
e^{-\alpha}\leq \frac{p(\neg x\mid y, s)}{p(\neg x\mid s )}\leq e^\alpha~ \qquad \text{for all $y{\in}\caly, x {\in}\calx$,  datasets $s$, and all priors $p(\cdot)$ on $\calx$} ~.
\end{equation}
This inequality \eqref{eq:BI3} 
means that the prior and posterior probabilities of the secret
taking a particular value 
do not differ too much. 
We now state precisely and prove the correspondence between DP and this Bayesian formulation. 

\begin{Theorem}
	Given a certain mechanism $C$:
	\begin{enumerate}
		\item If $C$  satisfies inequality \eqref{eq:DPcentral}, then it satisfies inequalities 
		\eqref{eq:BI3} and \eqref{eq:BI4}
		with $\alpha=\epsilon$.
		\item If $C$ satisfies inequalities \eqref{eq:BI3}  and \eqref{eq:BI4},
		then it satisfies inequality \eqref{eq:DPcentral} with $\epsilon=2\,\alpha$.
	\end{enumerate}
\end{Theorem}
The proof is analogous to that of Theorem~\ref{theo:correspondance};
we just need to replace $p(x), p(y), p(x \mid y)$ and  $p(y\mid x)$ by $p(x\mid s ), p(y\mid s ), p(x \mid y,s)$ and  $p(y\mid x, s)$ respectively, and similarly for $\neg x$. 
Also, we use the fact that $x$ and $\neg x$ are complementary events, i.e., $p(x)$ + $p(\neg x) \,=\,1$, and the same holds for the conditional probabilities on $x$ and $\neg x$.

\end{document}